\documentclass{ws-ijmpa}
\usepackage[super,compress]{cite}

\usepackage{multirow}
\usepackage{epsfig}
\usepackage{amssymb}
\usepackage{amsmath}

\begin{document}

\markboth{M. J. Menon \& P. V. R. G. Silva}
{An updated analysis on the rise of the hadronic total cross-section at the LHC energy region}

%
\catchline{}{}{}{}{}
%


\title{AN UPDATED ANALYSIS ON THE RISE OF THE HADRONIC TOTAL CROSS-SECTION AT THE LHC ENERGY REGION}

\author{M. J. MENON and P. V. R. G. SILVA}

\address{Universidade Estadual de Campinas - UNICAMP\\
Instituto de F\'{\i}sica Gleb Wataghin \\
13083-859 Campinas, SP, Brazil \\
(menon@ifi.unicamp.br, precchia@ifi.unicamp.br)}

\maketitle


\begin{abstract}
A forward amplitude analysis on $pp$ and $\bar{p}p$ elastic scattering
above 5 GeV is presented. The dataset includes the recent high-precision TOTEM 
measurements of the $pp$ total and elastic (integrated) cross-sections at 7 TeV and 8 TeV. 
Following previous works, 
the leading
high-energy contribution for the total cross-section ($\sigma_{tot}$) is parametrized as
$\ln^{\gamma}(s/s_h)$, where 
$\gamma$ and $s_h$ are free \textit{real} fit parameters. Singly-subtracted derivative 
dispersion relations are used to connect 
$\sigma_{tot}$ and the rho parameter ($\rho$) in an analytical way. Different fit
procedures are considered, including individual fits to $\sigma_{tot}$ data, 
global fits to $\sigma_{tot}$ and $\rho$ data, constrained and unconstrained
data reductions. The results favor
a rise  of the $\sigma_{tot}$ faster than the log-squared bound by 
Froissart and Martin at the LHC energy region. The parametrization for $\sigma_{tot}$
is extended to fit the elastic cross-section ($\sigma_{el}$)
data with satisfactory results. The analysis indicates an asymptotic ratio
$\sigma_{el}/\sigma_{tot}$ consistent with 1/3 (as already obtained in a previous work).
A critical discussion
on the correlation, practical role and physical implications of the parameters $\gamma$
and $s_h$ is presented. The discussion confronts the 2002 prediction of $\sigma_{tot}$
by the COMPETE Collaboration and the recent result by the Particle Data Group (2012 
edition of the Review of Particle Physics). Some conjectures on 
possible implications of
a fast rise of the proton-proton total cross-section at the highest
energies are also presented.


\end{abstract}

\ccode {PACS numbers: 13.85.-t, 13.85.Lg, 11.10.Jj}

\section{Introduction}
\label{s1}

The theoretical and phenomenological  descriptions of the energy dependence of 
the hadronic total cross-section at high energies have been a 
fundamental and long-standing
problem. Given that the  optical theorem connects the total cross-section with the
imaginary part of the forward elastic scattering amplitude, the main
theoretical difficulty concerns the lack of a pure (model-independent) nonperturbative QCD description of the 
soft scattering states, in particular the elastic channel in the forward direction.
In this section, we shortly recall some results on the rise
of the total cross-section of interest in this paper,
followed by the plan of the paper.

Historical formal analyses by Froissart, Lukaszuk and Martin \cite{froissart, martin1, martin2, lukamart} 
have established the famous asymptotic bound for the rise of the total cross-section,
\begin{eqnarray}
\sigma_{tot} (s) \leq c\ln^2 \frac{s}{s_0},
\nonumber
\end{eqnarray}
where $s$ is the center-of-mass energy squared, $c \leq \pi/m_\pi^2 \approx 60\, $ mb and $s_0$
is a \textit{constant}. After the Martin derivation in the context of the axiomatic quantum
field theory \cite{martin1}, the log-squared bound has played a central role in
model constructions, aimed to treat, interpret and describe  strong interactions.
Recently, Azimov has argued that it is not obvious that QCD can be considered an axiomatic theory
and has also established new constraints for the rate of increase
of $\sigma_{tot}(s)$ \cite{azimov1}. The formalism allows to conclude
that, depending on the behavior of the amplitude in the nonphysical
region, the total hadronic cross-section may rise faster than the log-squared bound,
without violation of unitarity \cite{azimov1}
(see also Refs. \citen{azimov2} and \citen{azimov3}).

The dependence of the total cross-section on the energy has been usually
investigated through forward amplitude analysis. This approach is characterized by different 
analytical parametrizations
for $\sigma_{tot}(s)$ and global data reductions including the $\rho$ parameter (ratio between the 
real and imaginary parts of the forward amplitude) by means of
singly-subtracted dispersion relations. In 2002, the COMPETE Collaboration  developed
a detailed analysis on different functional forms, using a ranking procedure and including
several reactions and different energy cutoffs \cite{compete1, compete2}.
The analysis favored the asymptotic form $\ln^2(s/s_0)$, when compared with
$\ln(s/s_0)$ or powers $[s/s_0]^{\epsilon}$ with $\epsilon >$ 0, a conclusion corroborated 
by subsequent works
\cite{ii1,ii2,bh1,bh2}. After that, data reductions with the COMPETE
highest-rank parametrization, in agreement with the
log-squared bound and all the experimental data available, became standard reference in the Review
of Particle Physics by the Particle Data Group (PDG) up to the 2010
edition \cite{pdg10}. 

In the experimental context, the recent results from the LHC, reaching the highest
energy region, opened  new expectations in the phenomenological and theoretical
contexts.
However, the first results on $pp$ elastic, diffractive and nondiffractive processes
by the TOTEM Collaboration have indicated  noticeable discrepancies with
standard predictions from representative phenomenological models \cite{totem1,totem2,totem3}.
In general grounds, this novel experimental information seems yet unable
to select or exclude classes of phenomenological
approaches and/or theoretical pictures in a clear and conclusive way.

In what concerns the $pp$ total cross-section, the first 7 TeV TOTEM high-precision measurement
has been obtained with small bunches and a luminosity-dependent method, indicating

\vspace{0.2cm}

\leftline{
$\sigma_{tot,1}$(7 TeV) = 98.3 $\pm$ 2.8\ mb
\
(small\ bunches\,/\,luminosity-dependent)\ \cite{totem2}.
}

\vspace{0.2cm}

In particular and for our purposes, the COMPETE \textit{prediction} (published 11 years ago) is in 
good agreement with this datum \cite{totem2}.
However, contrasting with this striking result, the 2012 edition of the Review
of Particle Physics by the PDG has shown
that with an updated dataset, including the above TOTEM datum, the fit with the  highest-rank
parametrization selected by the COMPETE Collaboration disagrees with the TOTEM result
(Figure 46.10 in Ref. \citen{pdg12}). Both results (COMPETE 2002 and PDG 2012) are displayed in 
Figure \ref{f1} and will be discussed along the paper.

In 2011 and 2012, Fagundes, Menon and Silva \cite{fms1,fms2} have revisited a parametrization
for the total cross-section introduced by Amaldi \textit{et al.} in the seventies \cite{amaldi}, 
characterized
by using the exponent in the high-energy leading logarithm contribution as a free real fit parameter.
Based on different fit procedures, variants and data ensembles from $pp$
and $\bar{p}p$ forward scattering at
$\sqrt{s} \geq$ 5 GeV, the authors have
shown that with the inclusion of the 7 TeV TOTEM result in the dataset, several 
statistically consistent solutions are
obtained with the exponent greater than 2, suggesting, therefore, a faster-than-squared-logarithm
rise for the total cross-section \cite{fms1,fms2}.

More recently, the TOTEM Collaboration has obtained four new high-precision measurements
for the total cross-section, with three data at 7 TeV  and one datum at 8 TeV
\cite{totem4,totem5,totem6}. The results and experimental
conditions can be summarized as follows:

\vspace{0.2cm}

\leftline{
$\sigma_{tot,2}$(7 TeV) = 98.6 $\pm$ 2.2\ mb
\
(large\ bunches\,/\,luminosity-dependent)\ \cite{totem4},
}

\vspace{0.2cm}

\leftline{
$\sigma_{tot,3}$(7 TeV) = 99.1 $\pm$ 4.3\ mb
\
($\rho$ independent)\ \cite{totem5},
}

\vspace{0.2cm}

\leftline{
$\sigma_{tot,4}$(7 TeV) = 98.0 $\pm$ 2.5\ mb
\
(luminosity-independent)\ \cite{totem5},
}

\vspace{0.2cm}

\leftline{
$\sigma_{tot}$(8 TeV) = 101.7 $\pm$ 2.9\ mb
\
(luminosity-independent)\ \cite{totem6}.
}

\vspace{0.2cm}

In this paper, we revisit, once more, the Amaldi \textit{et al.} parametrization, now taking into
account these recent measurements of the $pp$ total cross-section at 7 TeV and 8 TeV. 
The general strategy follows the approach developed in our previous analysis 
\cite{fms1,fms2}. In addition to using the new data, we will also explore the practical role 
and the physical meaning of the free fit parameters associated with the high-energy leading contribution.
In particular, we discuss the differences between the 2002 COMPETE prediction
and the 2012 PDG result in the phenomenological context (Reggeon/Pomeron exchanges).
As in Refs. \citen{fms1} and \citen{fms2}, the analysis is also restricted to $pp$ and $\bar{p}p$ data at $\sqrt{s} \geq$ 5 GeV,
the same energy cutoff used by the COMPETE Collaboration \cite{compete1,compete2}
and in the last PDG versions. \cite{pdg10,pdg12}
Our analysis and results favor, once more, a rise of the hadronic total cross-section faster than the
log-squared bound at the LHC energy region.
An extension of the parametrization to fit the elastic cross-section data allows us to infer
the asymptotic ratio between the elastic and total cross-sections. The result is
statistically consistent with a rational limit of 1/3 (as previously obtained in Ref. \citen{fms2}).
Some comments and conjectures on the possible implications of a
faster-than-squared-logarithm rise for the total cross-section are also presented. 

The paper is organized as follows. In section \ref{s2}, we display the analytical 
parametrization for $\sigma_{tot}(s)$ and the formula connecting this quantity with 
$\rho(s)$, using singly-subtracted \textit{derivative} dispersion relations \cite{fms1};
the fit procedures and strategies are also outlined. In section \ref{s3}, we present
the fit results with datasets up to $\sqrt{s}_{max}$ = 7 TeV 
and $\sqrt{s}_{max}$ = 8 TeV, in the cases of unconstrained and
constrained data reductions, individual fits to $\sigma_{tot}$
data and global fits to $\sigma_{tot}$ and $\rho$ data. A discussion on all the obtained results, with focus
on the high-energy leading contribution parameters, is presented in section \ref{s4}.
The extension to the elastic cross-section data is treated in section \ref{s5} and
discussions on the possibility of a remarkable fast rise of the total cross-section at the LHC 
energy region are presented
in section \ref{s6}. 
The conclusions and some final remarks are the contents of section \ref{s7}. 

\section{Analytical Parametrization and Fit Procedures}
\label{s2}

In this section, after introducing the formulas to be employed in
our data reductions, namely the parametrization for $\sigma_{tot}(s)$ and the analytical
expression for $\rho(s)$ from derivative dispersion relations, we outline
some important points on the fit procedures and strategies.

\subsection{Analytical parametrization and dispersion relation result}
\label{s21}

We consider the analytical parametrization for the $pp$ and $\bar{p}p$ total cross-section
introduced by Amaldi \textit{et al.} in 1970s \cite{amaldi} and
also employed by the UA4/2 Collaboration in 1990s. \cite{ua42}
It consists of two components,
associated with low-energy ($LE$) and high-energy ($HE$) contributions:
\begin{equation}
\sigma_{tot}(s) = \sigma_{LE}(s) +  \sigma_{HE}(s).
\end{equation}
The first term accounts for the decrease
of the total cross-section and
the differences
between $pp$  and $\bar{p}p$ scattering at low energies. In this
paper it is expressed by
\begin{equation}
\sigma_{LE}(s) = a_1\, \left[\frac{s}{s_l}\right]^{-b_1} + 
\tau\, a_2\, \left[\frac{s}{s_l}\right]^{-b_2},
\end{equation}
where $\tau$ = -1 (+1) for $pp$ ($\bar{p}p$) scattering, $s_l$ = 1 GeV$^2$ is fixed,
and $a_1$, $b_1$, $a_2$ and $b_2$ are free fit parameters.

The  second term accounts for the rising of the cross-section at higher energies and is given by
\begin{equation}
\sigma_{HE}(s) =
\alpha\, + \beta\, \ln^\gamma \frac{s}{s_h},
\end{equation}
where $\alpha$, $\beta$, $\gamma$ and $s_h$ are free real parameters.

For further reference, we briefly recall that, in the context of the Regge-Gribov theory,
the \textit{decreasing} $\sigma_{LE}(s)$ contribution is associated with Reggeon exchanges:
$b_1$ and $b_2$ correspond to the intercept of the trajectories and
$a_1$, $a_2$ to the Reggeon strengths (residues). The $\sigma_{HE}(s)$ term
simulates the \textit{rise} of the total cross-section and is associated with the
Pomeron exchange. For example, for $\gamma$ = 1, the constant plus $\ln s$ terms correspond 
to a double pole at $J$ = 1
and for $\gamma$ = 2 a triple pole (expressed by $\ln^2 s$, $\ln s$ and the constant terms)
\cite{compete1}.
For our purposes it is important to note that for $\gamma$ = 2, parametrization (Eqs. (1)-(3))
has the same analytical structure of the highest-rank parametrization
selected by the COMPETE Collaboration \cite{compete1,compete2}.

The analytical connection with the $\rho$ parameter is obtained  using singly-subtracted
derivative dispersion relations in the operator expansion form introduced by
Kang and Nicolescu \cite{kn} (also discussed in Ref. \citen{am04}). In terms of
the parametrization (1)-(3) for $pp$ and $\bar{p}p$
scattering,  the analytical results for $\rho(s)$ read \cite{fms1,fms2}

\begin{eqnarray}
\rho(s) &=& \frac{1}{\sigma_{tot}(s)}
\left\{ \frac{K}{s} 
- a_1\, \tan \left( \frac{\pi\, b_1}{2}\right) \left[\frac{s}{s_l}\right]^{-b_1} +
\mathcal{A}\, \ln^{\gamma - 1} \left(\frac{s}{s_h}\right) \nonumber  \right. \\
&+&
\mathcal{B}\, \ln^{\gamma - 3} \left(\frac{s}{s_h}\right)  
+ 
\left. 
\mathcal{C}\, \ln^{\gamma - 5} \left(\frac{s}{s_h}\right) 
+ \tau \, a_2\, \cot \left(\frac{\pi\, b_2}{2}\right) \left[\frac{s}{s_l}\right]^{-b_2} \right\},
\end{eqnarray}
where $K$ is the subtraction constant,
\begin{eqnarray} 
\mathcal{A} = \frac{\pi}{2} \, \beta\, \gamma,  
\quad 
\mathcal{B} = \frac{1}{3} \left[\frac{\pi}{2}\right]^3 \, \beta\, \gamma\, [\gamma - 1][ \gamma - 2], 
 \nonumber \\
\mathcal{C} = \frac{2}{15} \left[\frac{\pi}{2}\right]^5 \, \beta\, \gamma\, [\gamma - 1][ \gamma - 2]
[\gamma - 3][ \gamma - 4],
\end{eqnarray} 
and, as before, $\tau$ = -1 (+1) for $pp$ ($\bar{p}p$) scattering.

\subsection{Fit procedures and strategies}
\label{s22}

The fit procedures and methodology, as well as the role  and applicability of the subtraction
constant in the derivative dispersion relation approach, have been discussed in detail in 
Refs. \citen{fms1} and \citen{fms2}. Here we recall and also outline some
important points of interest in this work, especially the introduction of a representation
for the parameter $s_h$.

\subsubsection{Dataset}
\label{s221}

Our goal is to investigate the rise of $\sigma_{tot}$ at the highest energy region
and its asymptotic behavior. For that reason, we limit the analysis to particle-particle and 
antiparticle-particle collisions corresponding only to the largest energy interval
with available data, namely $pp$ and $\bar{p}p$ scattering. Although somewhat restrictive,
the main point is that this choice allows the investigation of possible high-energy effects 
that may be unrelated
to the trends of the lower energy data on other reactions
(as the constraints dictated by a supposed universal behavior).

The input dataset for fits concerns \textit{only accelerator data}
on $\sigma_{tot}$ and $\rho$ from $pp$ and $\bar{p}p$ scattering, covering
the energy region from 5 GeV up to 8 TeV. The energy cutoff is the same
employed in the COMPETE and PDG analyses \cite{compete1,pdg12}. The data below 
7 TeV have been 
collected from the PDG database \cite{pdg12}, without any kind of data selection
or sieve procedure. Statistical
and systematic errors have been added in quadrature.
Estimations
of the $pp$ total cross-section from cosmic-ray experiments will be
displayed in the figures as illustrative results. The TOTEM estimation
for $\rho$ at 7 TeV \cite{totem5} is also displayed as illustration.
All the references on these data
and estimations can be found in Ref. \citen{fms2}.

\subsubsection{Nonlinearity and feedback values}
\label{s222}

Because the nonlinearity of the fit  demands
a choice for the initial values (feedbacks) for all free parameters \cite{bev},
different choices have been tested and discussed in our previous analyses \cite{fms1,fms2}
(see also Ref. \citen{reply}).
Here, to initialize our parametric set, we consider only the values of the fit results
in the updated 2012 PDG version, obtained with the highest-rank COMPETE
parametrization \cite{pdg12}. 

This choice is based on the arguments that follow.
The PDG data reductions have been developed with $\gamma$ fixed to 2.
Due to the strong correlation among all the fit parameters
(to be discussed in Subsec. \ref{s41}), their final fit values
are, therefore, consequences of this condition ($\gamma$ = 2). 
In this sense,
initializing the parametrization with these values can be considered a
``conservative'' choice. Moreover,
since in the present analysis $\gamma$ is treated as a free parameter in fits including the recent TOTEM measurements,
this choice allows us to investigate possible departures from the standard/canonical
assumption $\gamma$ = 2.

However, it should be noted
that the PDG and COMPETE analyses include different collision processes
(meson-$p$, baryon-$p$, among others) and also tests on universality. Therefore
our dataset, restricted to $pp$ and $\bar{p}p$ scattering, corresponds to only a subset
of the ensemble employed in the global analysis by the COMPETE and PDG.
Despite this, we understand that with this choice for the feedbacks,
we initialize our parametrization with a statistically and physically meaningful input, contributing
to the search for a consistent fit solution (see also Subsec. \ref{s224} below). 

The values of the feedback parameters to be used in our fits are
shown in the third column of Table \ref{t1}.
For further reference, we display in Table \ref{t1} the values of the parameters
obtained with the highest-rank parametrization (Eqs. (1)-(3) with $\gamma$ = 2)
in both the 2002 COMPETE analysis (Table VIII in Ref. \citen{compete1}) and the
2012 PDG version (Table 46.2 in Ref. \citen{pdg12}), which is based on an updated dataset
including the first 7 TeV TOTEM datum. In the last case, the values of the parameters
$a_1$ and $a_2$ correspond to our normalization of Eq. (2), namely
$s_l$ = 1 GeV$^2$ fixed (which is different from the normalization adopted in Ref. \citen{pdg12}).
The corresponding results for $\sigma_{tot}(s)$
and uncertainty regions (evaluated through propagation from the errors in
Table \ref{t1}) are shown in Figure \ref{f1}, together with the experimental information
(in this figure only the first 7 TeV TOTEM measurement is displayed, as in Ref. \citen{pdg12}).
We shall return to these results along the paper.

\begin{table}[ht]
\tbl{Fit results through parametrization (1-3) with $\gamma$ = 2 obtained in the COMPETE and PDG analyses \cite{compete2,pdg12}. The parameters $a_1$, $a_2$, $\alpha$ and $\beta$ are in mb, $s_h$ in GeV$^{2}$, 
$b_1$, $b_2$ are dimensionless ($s_l$ = 1 GeV$^2$).}
{\begin{tabular}{c c c}\hline
           & \qquad COMPETE 2002 (Ref. \citen{compete2}) & \qquad PDG 2012 (Ref. \citen{pdg12})   \\
\hline
$a_1$     & 42.1   $\pm$ 1.3         & 46.07  $\pm$ 0.76  \\
$b_1$     & 0.467  $\pm$ 0.015       & 0.462  $\pm$ 0.002   \\
$a_2$     & 32.19  $\pm$ 0.94        & 34.02  $\pm$ 0.63   \\
$b_2$     & 0.5398 $\pm$ 0.0064      & 0.550  $\pm$ 0.005   \\
$\alpha$  & 35.83  $\pm$ 0.40        & 34.71  $\pm$ 0.15    \\
$\beta$   & 0.3152 $\pm$ 0.0095      & 0.265  $\pm$ 0.050   \\
$s_h$     & 34.0   $\pm$ 0.54        & 16.21  $\pm$ 0.16    \\
\hline
\end{tabular} \label{t1}}
\end{table}

\subsubsection{Individual and global fits}
\label{s223}

Global fits to $\sigma_{tot}$ and $\rho$ data demand the use of dispersion relations
with one subtraction and therefore the introduction of one more free parameter, the subtraction constant. 
As already discussed in Refs. \citen{fms1} and \citen{fms2} this parameter does not have
a physical interpretation as opposed to the parameters present in the total cross-section parametrization,
which are associated with Reggeon and Pomeron exchanges. Moreover, several authors have criticized the
usual methods to extract $\rho$ (see Subsec. 2.2 in Ref. \citen{fms1}) and, in addition, due to
the correlation among all the fit parameters, the
$\rho$ inclusion in global fit constraints the rise of the total cross-section, as demonstrated in Refs.
\citen{fms2,alm01,alm03} and references therein.
Despite these disadvantages, 
we shall here consider both individual fits to $\sigma_{tot}$ data
through Eqs. (1)-(3) and global fits to  $\sigma_{tot}$ and $\rho$ data using Eqs. (1)-(5).

\subsubsection{Minimization and statistics}
\label{s224}
 
The data reductions have been performed with the objects of the class TMinuit of the ROOT Framework 
\cite{root}. We have employed the default MINUIT error analysis \cite{minuit} with the 
\textit{selective criteria} that follow.
In the minimization program a Confidence Level of one standard deviation
was adopted in all fits (UP = 1). In each test of fit, successive running of the MIGRAD
have been considered (up to 5,000 calls), 
until full convergence has been reached, with the smallest FCN (chi-squared)
and requiring Estimated Distance to Minimum (EDM) $< 10^{-4}$, adequate for the 
one sigma CL. In some cases the MINOS algorithm and strategies 1 and 2 have also been
employed to check the MIGRAD result.
In addition, the error in the parameters should not exceed the central value. 
The error matrix provides the variances and covariances associated with each free parameter,
which are used in the analytic evaluation of the uncertainty regions in the fitted and predicted
quantities, by means of standard error propagation procedures \cite{bev}.

To quantify goodness of fit we will resort to chi-square per degree of freedom
($\chi^2$/DOF) and
the corresponding integrated probability, $P(\chi^2)$ \cite{bev}. The goal is not to compare or select 
fit procedures or fit results but only to check the statistical consistence of the 
data reductions in a reasonable way.

\subsubsection{Unconstrained and constrained fits}
\label{s225}

The consideration that the exponent $\gamma$ in the leading logarithm component is a real 
(not integer) free fit parameter leads to some special consequences and 
particular conditions.
These aspects, including 
the effects of the correlation
between the free parameters $\gamma$ and $s_h$ in data reductions,
will be discussed
in some detail in Sec. \ref{s4}, after the presentation of our fit results.
In order to implement and facilitate that
discussion, we introduce here a useful representation for the high-energy scaling factor $s_h$,
which will lead us to distinguish between unconstrained and constrained fits, as explained
in what follows. 

The representation is based on two arguments: (i) the reference to $s_h$ only as a (unknown)
constant in the Froissart-Martin
derivation of the bound; (ii) the reasonable physical conjecture that this factor might 
be proportional to the energy threshold for the scattering states (above the resonance region),
namely $s_h \propto (m_p + m_p)^2$, where $m_p$ is the proton mass.
In this case, we can represent the scaling factor by
\begin{equation}
s_h = \delta\, [\,4m_p^ 2\,],
\qquad
4m_p^ 2 = \textrm{3.521 GeV}^2,
\end{equation}
where $\delta$ is a real dimensionless parameter and
$\delta$ = 1 at the physical threshold. 

With this representation, we can distinguish two physical conditions in data reductions: either
to consider $\delta$ indeed as a free fit parameter (equivalently, $s_h$ a free fit parameter)
or to assume $\delta$ = 1 (equivalently, to fix $s_h = 4m_p^2$, the energy threshold). 
In what follows we shall denote these two variants by \textit{unconstrained fits} ($\delta$ free) and 
\textit{constrained fits} ($\delta$ = 1 fixed).

Here, as in the COMPETE and PDG analyses, we shall treat $s_h$ as a constant (with the
above representation). However, it should be noted that the possibility of a slow rise of
$s_h$ with $s$ (for large $s$) has been discussed by some authors, as for example
in Refs. \citen{azimov1} and \citen{common}.

\subsubsection{Ensembles and feedbacks}
\label{s226}

As mentioned earlier, our dataset consists of $pp$ and $\bar{p}p$ accelerator data at
$\sqrt{s} \geq$ 5 GeV. In order to investigate the effect in the fits associated with the
TOTEM measurement at 8 TeV, as compared with those at 7 TeV, we shall
consider two data ensembles. The first one with data up to 7 TeV (including the four measurements)
and a second one adding the 8 TeV datum. For reference, we will denote these two variants by
$\sqrt{s}_{max}$ = 7 TeV ensemble and $\sqrt{s}_{max}$ = 8 TeV ensemble, respectively.

For the $\sqrt{s}_{max}$ = 7 TeV ensemble, we use as feedback the values of the parameters
from the 2012 PDG version, displayed in the third column of Table \ref{t1}. In the case
of global fits to $\sigma_{tot}$ and $\rho$ data, we consider $K$ = 0 for the initial
value of the subtraction constant \cite{pdg12}. The fit results with this ensemble
are then used as feedback to initialize the parametrization with the 
$\sqrt{s}_{max}$ = 8 TeV ensemble, in each one of the variants considered (individual, global,
unconstrained and constrained cases).

\section{Fit Results}
\label{s3}

Summarizing the variants discussed in the last section, we select two ensembles of
accelerator data on $\sigma_{tot}$ and $\rho$ at $\sqrt{s} \geq$ 5 GeV, one up to 
$\sqrt{s}_{max}$ = 7 TeV and  another one up to $\sqrt{s}_{max}$ = 8 TeV. In each case
we consider both unconstrained fits ($s_h$ or $\delta$ in Eq.(6) as free fit parameter)
and constrained fits ($s_h = 4m_p^2$ or $\delta$ = 1 fixed), treating also both
individual fits to $\sigma_{tot}$ data through Eqs. (1)-(3) and 
global fits to $\sigma_{tot}$ and $\rho$ data, using Eqs. (1)-(5).
In what follows we present the fit results and discuss only the accordance with
the selective criteria outlined in Subsec. \ref{s224}. The physical aspects
and implications involved are addressed in Sec. \ref{s4}.

\subsection{$\sqrt{s}_{max}$ = 7 TeV ensemble}
\label{s31}

For this ensemble all data reductions presented satisfactory agreement with the selective criteria.
The fit results and statistical information are displayed in Table \ref{t2}.
The curves, uncertainty regions (from error propagation) and experimental information 
in the case of \textit{global} fits to $\sigma_{tot}$ and $\rho$ data are shown in
Fig. \ref{f2} for the unconstrained fit and in Fig. \ref{f3} for the
constrained one.

\begin{table}[ht]
\tbl{$\sqrt{s}_{max}$ = 7 TeV Ensemble. Fit results using Equations (1)-(3)
for the $\sigma_{tot}$ data and Equations (1)-(5) for  the $\sigma_{tot}$ and $\rho$
data in the case of unconstrained ($s_h$ free) and constrained ($s_h = 4m_p^2$) data reductions.
Units as in Table \ref{t1}.}
{\begin{tabular}{c c c c c}
\hline
            & \multicolumn{2}{c}{Unconstrained ($\delta$ free)}   &  
              \multicolumn{2}{c}{Constrained ($\delta$ = 1)}   \\
            & $\sigma_{tot}$  & $\sigma_{tot}$ and $\rho$  & $\sigma_{tot}$  & $\sigma_{tot}$ and $\rho$ \\
\hline
$a_1$       &59.5 $\pm$ 8.8          &56.7 $\pm$ 6.4 & 60.5 $\pm$ 7.8 & 59.1 $\pm$ 4.9\\
$b_1$       &0.553 $\pm$ 0.063       &0.528 $\pm$ 0.062 & 0.526 $\pm$ 0.074         & 0.500 $\pm$ 0.054\\
$a_2$       &33.2 $\pm$ 2.2          &34.0 $\pm$ 2.0 & 33.2 $\pm$ 2.3            & 34.1 $\pm$ 1.9\\
$b_2$       &0.541 $\pm$ 0.015       &0.547 $\pm$ 0.013 & 0.541 $\pm$ 0.016         & 0.547 $\pm$ 0.013\\
$\alpha$    &35.53$\pm$ 0.70         &35.3$\pm$ 1.0 & 34.0 $\pm$ 1.6            & 33.2 $\pm$ 1.3\\
$\beta$     &0.184 $\pm$ 0.066       &0.242 $\pm$ 0.064 &0.106 $\pm$ 0.044         & 0.130 $\pm$ 0.042\\
$\gamma$    &\textbf{2.14 $\pm$ 0.12}&\textbf{2.053 $\pm$ 0.094}&\textbf{2.29 $\pm$ 0.16}&\textbf{2.21 $\pm$ 0.11}\\
$s_h$       &12.8 $\pm$ 3.0          &16.41 $\pm$ 0.64 & 3.521 (fixed)            & 3.521 (fixed)\\
$K$         & -                      &40 $\pm$ 14 & -                         & 39 $\pm$ 12\\
\hline
DOF         & 159                    &234 & 160                       & 235\\
$\chi^2$/DOF& 0.92                   &1.09 & 0.91                      & 1.09\\
$P(\chi^2)$ & 0.754                  &0.155 & 0.775                     & 0.164\\
\hline
Figure:     &        -               &   2   & -  &  3 \\
\hline 
 \end{tabular}
\label{t2}}
\end{table}

\subsection{$\sqrt{s}_{max}$ = 8 TeV ensemble}
\label{s32}

For this ensemble, the unconstrained fits presented some disagreement with the selective criteria.
In the case of global data reduction to $\sigma_{tot}$ and $\rho$ the fit did not converge
and therefore we have no solution for this case. 
In the
individual fit to $\sigma_{tot}$ the data reduction converged but it should be noted
that the corresponding error matrix is not positive definite. 

The fit results and statistical information are displayed in Table \ref{t3}.
The curves, uncertainty regions and experimental information in the case
of the constrained \textit{global} fit to $\sigma_{tot}$ and $\rho$ data are shown in Fig. \ref{f4}
(the unconstrained case did not converge).

\begin{table}[ht]
\tbl{$\sqrt{s}_{max}$ = 8 TeV Ensemble. Fit results using Eqs. (1)-(3)
for the $\sigma_{tot}$ data (unconstrained, $s_h$ free, and constrained, $s_h = 4m_p^2$)
and Eqs. (1)-(5) for  the $\sigma_{tot}$ and $\rho$ data (only the constrained case).
Units as in Table \ref{t1}.}
 {\begin{tabular}{c c c c c}\hline
            & \multicolumn{2}{c}{Unconstrained ($\delta$ free)}   &  \multicolumn{2}{c}{Constrained ($\delta$ = 1)} \\
            & $\sigma_{tot}$  & $\sigma_{tot}$ and $\rho$  & $\sigma_{tot}$  & $\sigma_{tot}$ and $\rho$ \\
\hline
$a_1$       & 61.4 $\pm$ 1.7           &-& 60.8 $\pm$ 6.9            & 59.5 $\pm$ 7.2\\
$b_1$       & 0.5062 $\pm$ 0.0084      &-& 0.530 $\pm$ 0.061         & 0.505 $\pm$ 0.075\\
$a_2$       & 33.2 $\pm$ 1.7           &-& 33.2 $\pm$ 2.3            & 34.1 $\pm$ 2.0\\
$b_2$       & 0.541 $\pm$ 0.012        &-& 0.541 $\pm$ 0.016         & 0.547 $\pm$ 0.013\\
$\alpha$    &  32.37 $\pm$ 0.22        &-& 34.1 $\pm$ 1.2            & 33.4 $\pm$ 1.8\\
$\beta$     & 0.0477 $\pm$ 0.0022      &-& 0.102 $\pm$ 0.033         & 0.124 $\pm$ 0.054\\
$\gamma$    &\textbf{2.493 $\pm$ 0.016}&-&\textbf{2.30 $\pm$ 0.11}   &\textbf{2.23 $\pm$ 0.15}\\
$s_h$       & 0.633 $\pm$ 0.073        &-& 3.521 (fixed)             & 3.521 (fixed)\\
$K$         & -                        &-& -                         & 40 $\pm$ 16\\
\hline
DOF         &  160                    &-& 161                        & 236\\
$\chi^2$/DOF&   0.91                  &-& 0.91                       & 1.09\\
$P(\chi^2)$ &  0.774                  &-& 0.788                      & 0.172\\
\hline
Figure:     &        5 (up)        &   -   & 5 (down)  &  4 \\
\hline 
 \end{tabular}
\label{t3}}
\end{table}

\section{Discussion}
\label{s4}

Using parametrizations (1)-(5),
we are interested in a consistent quantitative description of the rise of the total cross-section at
high energies. We have considered 8 variants of data reductions, obtaining full convergence
in 7 cases. The fit results are displayed in Tables \ref{t2} and \ref{t3}. Our main goal
is to investigate if these results indicate a log-squared behavior or a rise faster
than this bound. In what follows, for a given numerical result $\gamma \pm \Delta \gamma$,
we consider a \textit{result statistically consistent with a faster rise} the cases in which
$\gamma - \Delta \gamma > $ 2. In this case, for short, we will refer to a result statistically consistent with
$\gamma >$ 2.
However, as remarked early in Subs. \ref{s225}, to consider the exponent $\gamma$ as a 
continuous real
 fit parameter has some special implications not present in the
canonical assumption $\gamma$ = 2 (fixed). These aspects are directly related to
the high-energy scaling factor $s_h$ and have important consequences not only on the data
reductions but also on the physical interpretation of the fit results. 
The goal of this section is to address these aspects.

To this end, in Subsec. \ref{s41} we discuss the individual and global fit results,
with focus on the value of the parameter $\gamma$ and its relation with a 
faster-than-squared-logarithm rise. That will lead us to our partial conclusions
in favor of this faster rise.
After that, in Subsec. \ref{s42} we address the practical role and physical 
implications associated with the correlation between $\gamma$ and $s_h$.
This discussion will lead us in Subsec. \ref{s43} to a final conclusion in favor
of the constrained fits in both physical and statistical
contexts, indicating a rise of the total cross-section faster than the log-squared bound.

\subsection{Individual and global fits: partial conclusions}
\label{s41}

Based on the results displayed in Tables \ref{t2} and \ref{t3}
we have the comments that follow.

In the case of \textit{individual} fits 
to $\sigma_{tot}$ data, all the results 
(constrained or unconstrained, $\sqrt{s}_{max}$ = 7 or 8 TeV ensembles) are statistically consistent
with $\gamma >$ 2 (confirming, therefore, the conclusions first presented
in Ref. \citen{fms1}).
In all cases the integrated probability reads
$P(\chi^2) \approx$ 0.8.

The highest $\gamma$-values are associated with the $\sqrt{s}_{max}$ = 8 TeV ensemble,
indicating $\gamma \approx$ 2.5 (unconstrained fit) and 
$\gamma \approx$ 2.3 (constrained fit). 
The corresponding results for $\sigma_{tot}(s)$ and uncertainty regions
are displayed in Fig. \ref{f5} for both
the unconstrained and constrained fits, together with the experimental data.
We notice that the agreement with the TOTEM measurements at 7 TeV is
striking. In the case of the constrained fit the uncertainty region includes the
central value at 8 TeV and also all the four 7 TeV central values.

The numerical results and predictions for the total cross-section at some energies of interest
are displayed in Table \ref{t4} (all the variants investigated
in the individual fits). We also note that with the  $\sqrt{s}_{max}$ = 8 TeV ensemble,
the predictions at 57 TeV, namely $\sigma_{tot} \sim$ 142 - 143 mb are about 7 \%
larger than the central value of the result by the Pierre Auger Collaboration,
namely 133 mb \cite{auger}.

\begin{table}[ht]
\tbl{Fit results and predictions for the $pp$ total cross-section at higher energies from 
\textit{individual} fits to $\sigma_{tot}$ data.}
{\begin{tabular}{c c c c c}\hline
             & \multicolumn{2}{c}{$\sqrt{s}_\mathrm{max} = 7$ TeV}   &
\multicolumn{2}{c}{$\sqrt{s}_\mathrm{max} = 8$ TeV}   \\
             & Unconstrained    & Constrained     & Unconstrained     &
Constrained       \\\hline
7 TeV        & 97.8 $\pm$ 1.1   & 97.9 $\pm$ 1.3  & 98.12 $\pm$ 0.86  & 98.1 $\pm$ 1.2 \\
8 TeV        & 100.2 $\pm$ 1.2  & 100.2 $\pm$ 1.4 & 100.56 $\pm$ 0.91 & 100.5 $\pm$ 1.3 \\
14 TeV       & 110.6 $\pm$ 1.6  & 110.8 $\pm$ 2.0 & 111.4 $\pm$ 1.1   & 111.2 $\pm$ 1.7  \\
57 TeV       & 140.6 $\pm$ 3.2  & 141.3 $\pm$ 4.4 & 142.7 $\pm$ 1.8   & 142.0 $\pm$ 3.5  \\
\hline
\end{tabular}
\label{t4}}
\end{table}

In the case of \textit{global} fits to $\sigma_{tot}$ and $\rho$ data, the results
depend on the ensemble and on the constraint condition considered.
In all cases of global convergent fits the integrated probability reads
$P(\chi^2) \approx$ 0.2.
For the constrained fits (both ensembles) the results are statistically
consistent with $\gamma >$ 2. In the unconstrained case and
$\sqrt{s}_{max}$ = 7 TeV ensemble the result may be considered barely consistent with
$\gamma >$ 2, since, up to 2 figures, $\gamma$ lies in the interval 2.0 - 2.2. 
As stated before, for the $\sqrt{s}_{max}$ = 8 TeV ensemble we did not obtain full convergence.

The numerical results and predictions for the total cross-section at the energies of interest
are displayed in Table \ref{t5}.
We note that at 57 TeV the results indicate $\sigma_{tot} \sim$ 139 - 140 mb,
which is about 5 \% larger than the Auger central value.

 \begin{table}[ht]
\tbl{Fit results and predictions for the $pp$ total cross-section at higher energies from 
\textit{global} fits to $\sigma_{tot}$ and $\rho$ data.}
{\begin{tabular}{c c c c c}\hline
             & \multicolumn{2}{c }{$\sqrt{s}_{max} = 7$ TeV}   & \multicolumn{2}{c}{$\sqrt{s}_{max} = 8$ TeV}   \\
             & Unconstrained    & Constrained     & Unconstrained   & Constrained     \\\hline
7 TeV        & 97.5 $\pm$ 1.2   & 97.6 $\pm$ 1.2  &    $-$          & 97.8 $\pm$ 1.3  \\
8 TeV        & 99.8 $\pm$ 1.3   & 99.9 $\pm$ 1.3  &    $-$          & 100.2 $\pm$ 1.4 \\
14 TeV       & 109.9 $\pm$ 1.7  & 110.2 $\pm$ 1.7 &    $-$          & 110.6 $\pm$ 2.0 \\
57 TeV       & 138.8 $\pm$ 3.3  & 139.6 $\pm$ 3.4 &    $-$          & 140.4 $\pm$ 4.3 \\
\hline
\end{tabular}
\label{t5}}
\end{table}

It is important to note that, in going from the individual to global fits, the constraint
imposed by the inclusion of the $\rho$ information on the rise of $\sigma_{tot}$ is
evident: in all cases the $\gamma$ value decreases (Tables \ref{t2} and \ref{t3}).
In this respect, the subtraction constant plays a remarkable role due to its
correlation with all the fit parameters in the nonlinear data reduction,
specially those associated with the high-energy contribution, namely
$\alpha$, $\beta$, $\gamma$ and $s_h$. This effect can be illustrated by the correlation 
matrix in the MINUIT Code, which provides
a measure of the correlation between each pair of free parameters
through a coefficient with numerical limits $\pm$ 1 (full correlation)
and 0 (no correlation) \cite{bev,root}. 
For example,
the coefficients in the global fits with the $\sqrt{s}_{max} = 7$ TeV ensemble,
in the cases of unconstrained fit (UF) and constrained fit (CF), are
displayed in Table \ref{t6}. In both cases the correlations between $K$
and $\alpha$, $\beta$ or $\gamma$ are around 0.8 - 0.9, affecting therefore
the asymptotic behavior of $\sigma_{tot}$. However, as already commented in
Subsec. \ref{s223} (and in more detail in Ref. \citen{fms2}) this important parameter does not 
have a physical interpretation as is the case for those present in the parametrization of $\sigma_{tot}(s)$.
We also stress that the integrated probabilities $P(\chi^2)$ in the global fits
are smaller than in the individual fits, $\sim 0.2$ and $\sim 0.8$, respectively.

\begin{table}[ht]
\tbl{Correlation coefficients from the correlation matrices  associated with 
unconstrained fit (UF) and constrained fit (CF) in the case
of global fits with the $\sqrt{s}_{max} = 7$ TeV ensemble.
The off-diagonal coefficients 
from the UF are displayed above the diagonal
of the table (not filled) and those from the CF, below that diagonal.}
{\begin{tabular}{c |c| c c c c c c c c c}\hline
 &  & \multicolumn{9}{c}{UF}\\\hline
 &          & $a_1$ & $b_1$ & $a_2$ & $b_2$ & $\alpha$ & $\beta$ & $\gamma$ & $s_h$ & $K$ \\\hline
\multirow{8}{*}{CF}
 &$a_1$   &      &0.981 &0.154&0.149&0.933&-0.904&0.878&-0.598&0.914\\
 &$b_1$   &0.966&      &0.092&0.094&0.985&-0.969&0.950&-0.539&0.909\\
 &$a_2$   &0.193&0.129&     &0.985&0.058&-0.052&0.054&0.183&-0.018\\
 &$b_2$   &0.178&0.122&0.985&   &0.063&-0.059&0.062&0.187&-0.038\\
 &$\alpha$&0.909&0.985&0.105&0.100&       &-0.997&0.988&-0.443&0.877\\
 &$\beta$ &-0.847&-0.949&-0.125&-0.122&-0.986&    &-0.996&0.389&-0.855\\
 &$\gamma$&0.825&0.933&0.126&0.123&0.976&-0.998&       &-0.320&0.833\\
 &$s_h$   &   - &  -  &   -   &   -   &   -   &   -   &   -   &      &-0.543\\
 &$K$     &0.882&0.872& 0.000&-0.032&0.835&-0.787&0.770&   -   &       \\
 \hline
 \end{tabular}
\label{t6}}
\end{table}

Despite the  aforementioned constraint, based on the  discussion concerning the $\gamma$ values
obtained in both individual and global fits, we conclude that our results favor a rise of
the hadronic total cross-section faster than the log-squared behavior at the LHC energy region.
By ``favor''\ we mean that, within the uncertainties, the fit results lead to $\gamma$ values
above 2 and not 2 or below 2.
The results for the $\gamma$ parameter from all the fully converged fits, within the
uncertainties (Tables \ref{t2} and \ref{t3}), are schematically displayed in Figure \ref{f6}.
Note that, in the case of constrained fits,
the $\gamma$ values lie in the interval 2.2 - 2.3,
which indicates more stability than in the unconstrained cases and
suggests, therefore, a support to the former variant. We shall return to this 
point in section \ref{s43}.

\subsection{The role and effects of the parameters $\gamma$ and $s_h$}
\label{s42}

From the example displayed in Table \ref{t6} we also notice a substantial negative correlation 
between the parameter $\gamma$ and the energy scaling parameter $s_h$. 
In fact, from Tables \ref{t2} and \ref{t3}, in going from the unconstrained
to the constrained case (and also from individual to global fits),
a decrease in $s_h$ is associated with an increase in $\gamma$ and vice versa.

\begin{figure}
\centering
\epsfig{file=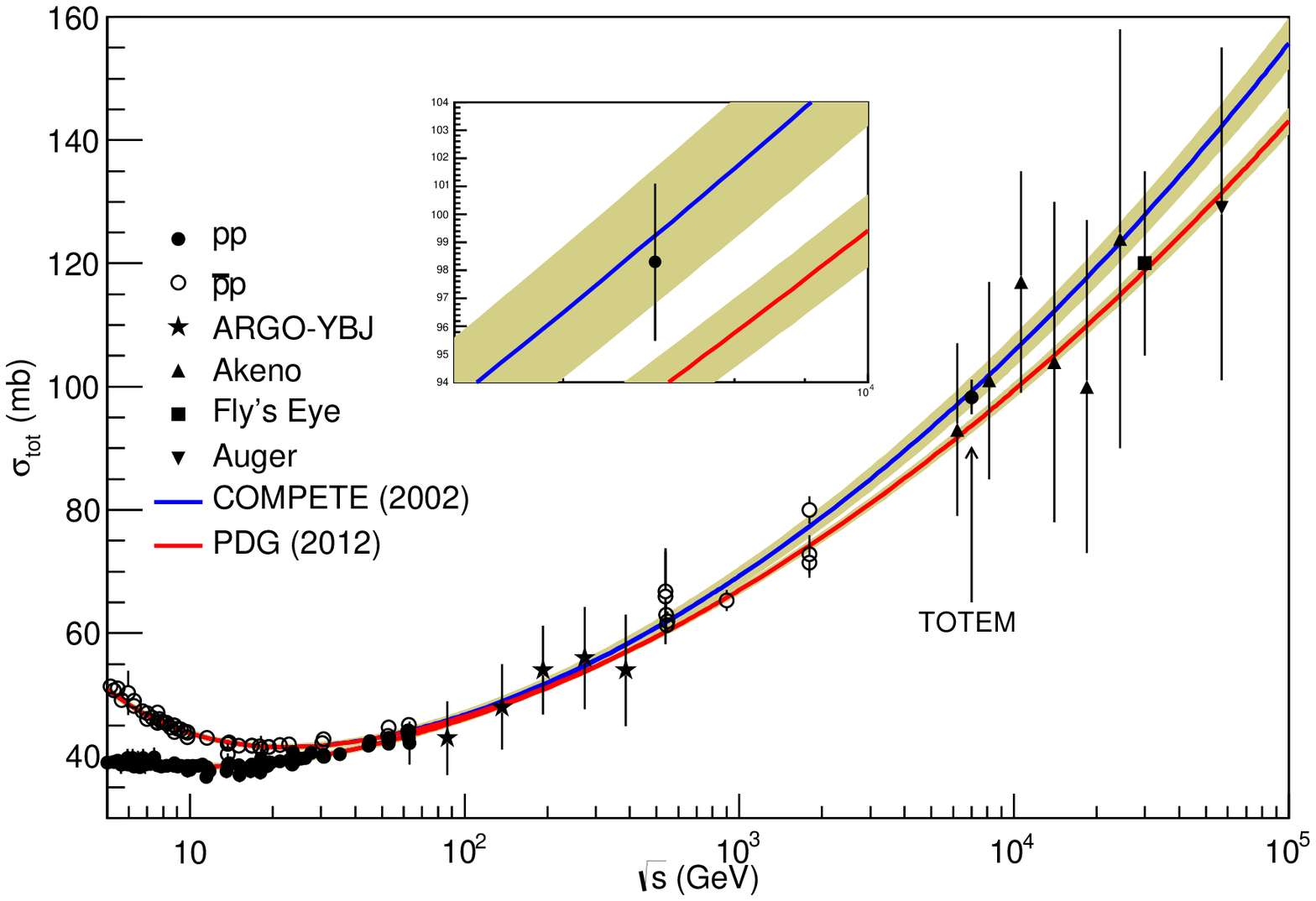,width=15cm,height=16cm}
\caption{Results with the COMPETE highest-rank parametrization
(Eqs. (1)-(3) with $\gamma$ = 2) from the 2002 COMPETE analysis \cite{compete1,compete2},
and from the 2012 PDG version \cite{pdg12} (which includes the first 7 TeV TOTEM datum).
The corresponding values of the parameters are displayed in Table \ref{t1}.}
\label{f1}
\end{figure}

\begin{figure}
\centering
\epsfig{file=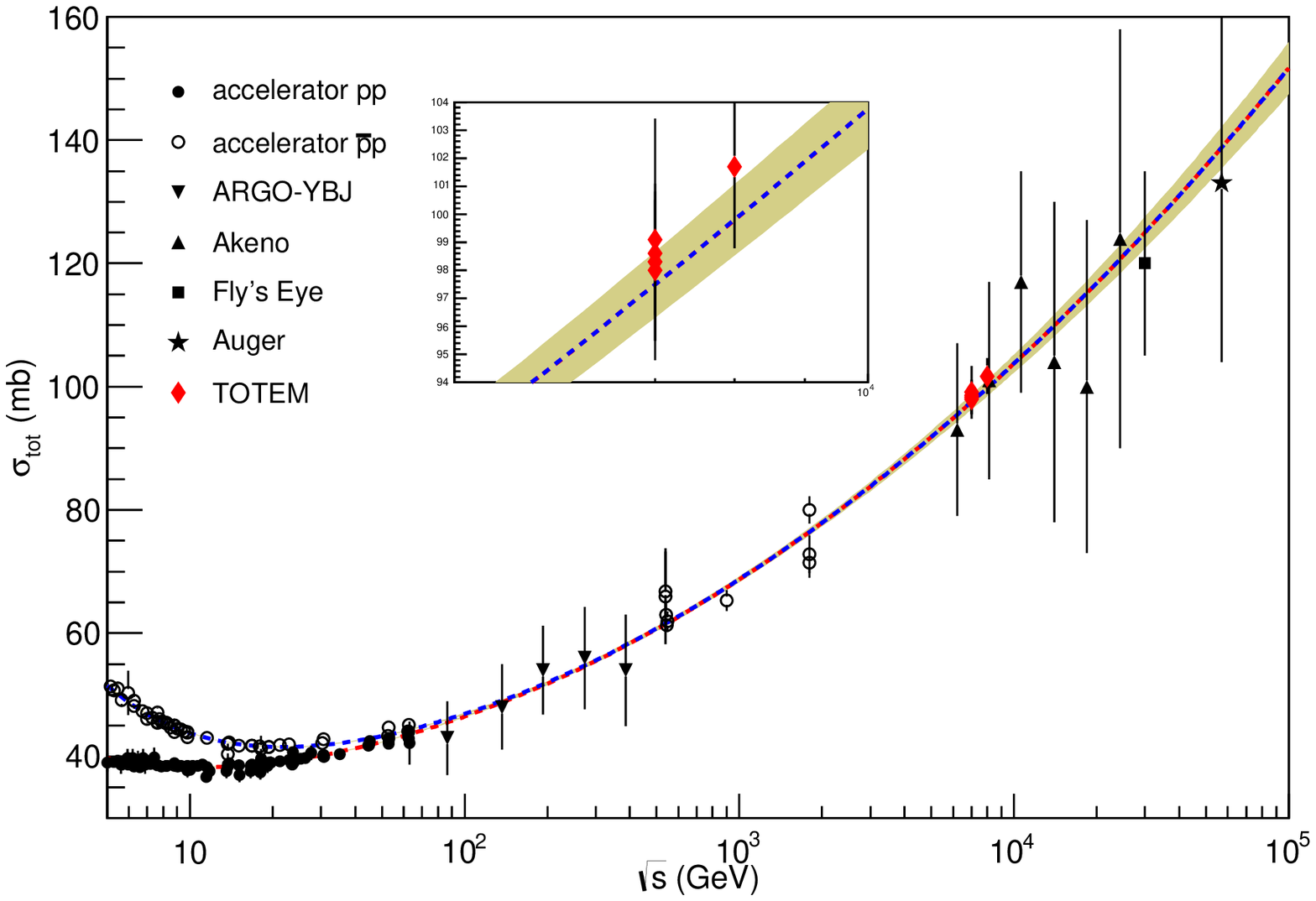,width=16cm,height=8.5cm}
\epsfig{file=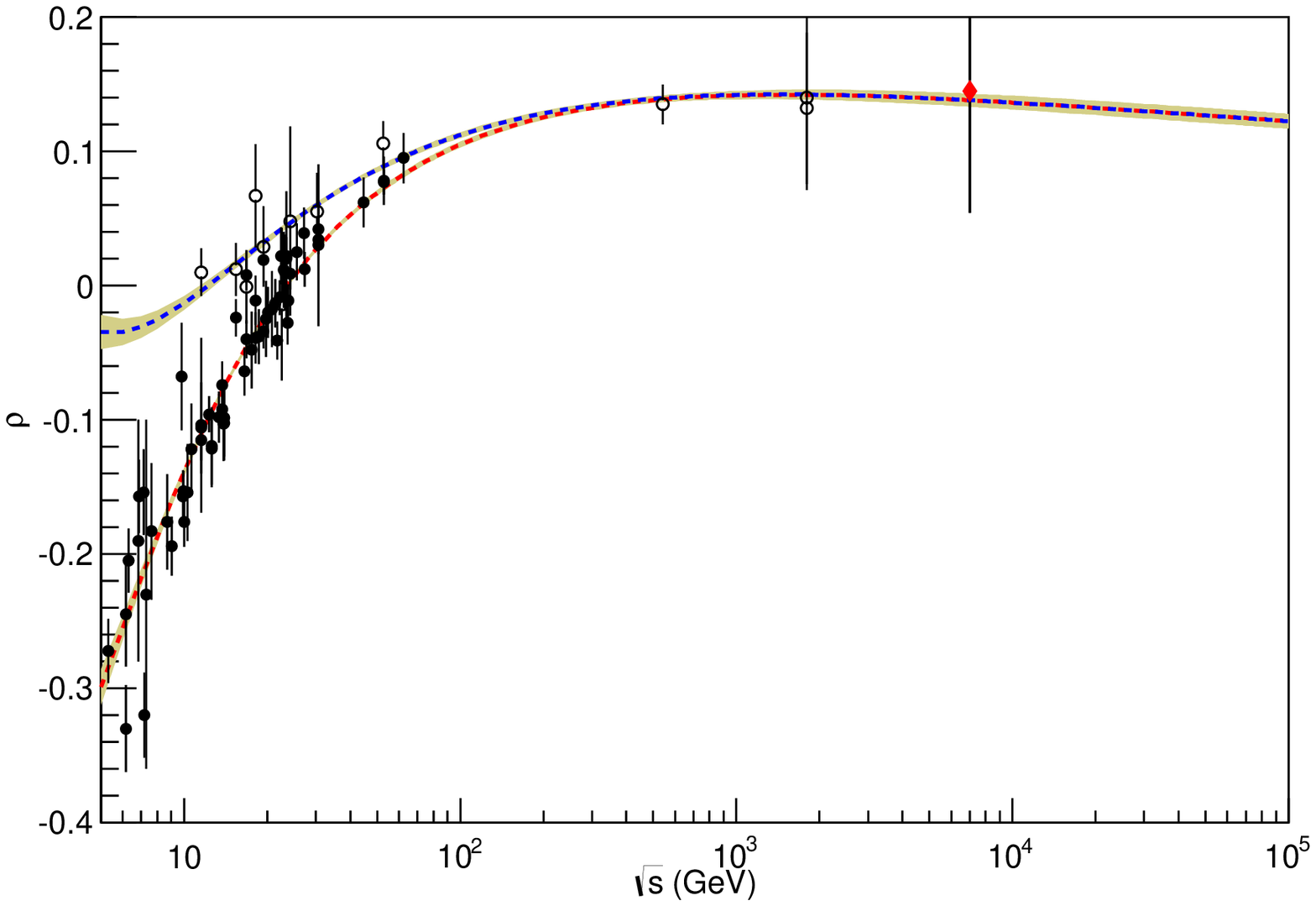,width=16cm,height=8.5cm}
\caption{Global unconstrained fit results with the $\sqrt{s}_{max}$ = 7 TeV ensemble 
(third column in Table \ref{t2}).}
\label{f2}
\end{figure}

\begin{figure}
\centering
\epsfig{file=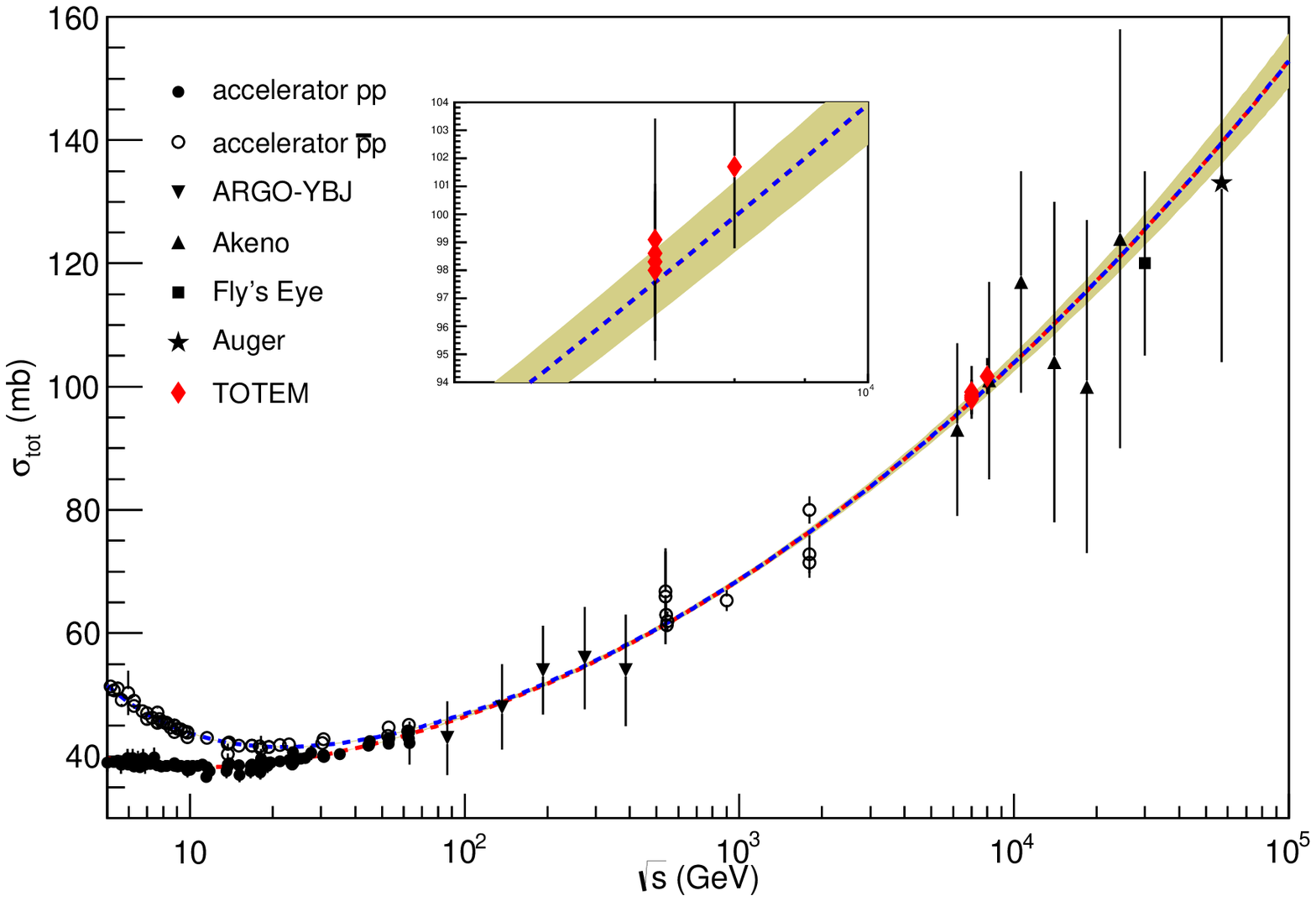,width=15cm,height=8.5cm}
\epsfig{file=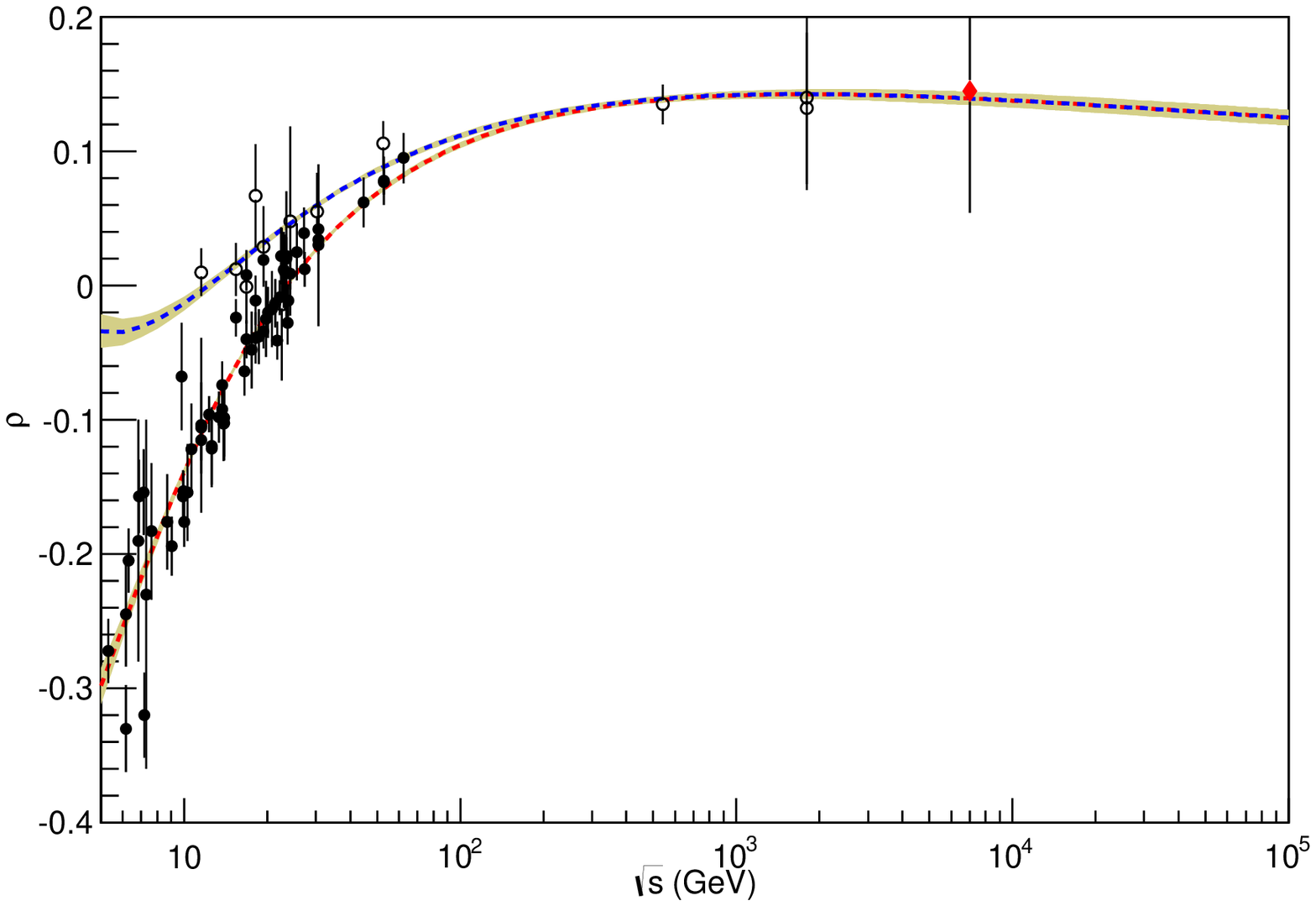,width=15cm,height=8.5cm}
\caption{Global constrained fit results with the $\sqrt{s}_{max}$ = 7 TeV ensemble
(fifth column in Table \ref{t2}).}
\label{f3}
\end{figure}

In this section, based on the previous results and discussion, we examine in some detail 
the important practical and physical role of the scale parameter $s_h$, especially in what 
concerns data reductions with $\gamma$ as a free \textit{real} (not integer) parameter.
In what follows, we first list five characteristics of interest involved, distinguishing the case
of $\gamma$ = 2, and then discuss the connections of these characteristics with our fit results,
as well as with the COMPETE 2002 and PDG 2012 results (Table \ref{t1}
and Figure \ref{f1}).

\begin{description}

\item{1.} 
Even treating $s_h$ as an unknown constant, with the canonical choice $\gamma$ = 2 the scaling factor 
can be well determined through the data reduction (since there is no correlation in this case).
However, with $\gamma$
real and free we have two unknown and anticorrelated parameters.
As a consequence, we can obtain statistically consistent solutions
for different values of these parameters and that may imply in different physical pictures
(as we shall show).

\item{2.} 
For $\gamma$ = 2 (fixed), the high-energy contribution $\sigma_{HE}(s)$, Eq. (3), is well
defined for all values of $s$ as compared with the $s_h$ value. The only difference concerns
the fact that for $s > s_h$ the logarithm term increases with the energy and for
$s < s_h$ this contribution decreases as the energy increases (we will return to this important
point in what follows). On the other hand, given that it represents a physical quantity, for $\gamma$ real 
(not integer) the logarithmic
term is not defined for $s < s_h$ so that this component can only start at $s = s_h$.
Above this point the contribution increases with the energy.

\item{3.} 
For $\gamma$ = 2  and $s_h = \delta\,[4m_p^2]$ the high-energy contribution can be written
\begin{equation}
\sigma_{HE}(s) =
a\, + b\, \ln\left(\frac{s}{4m_p^2}\right) + \beta \ln^2 \left(\frac{s}{4m_p^2}\right),
\end{equation}
where $a = \alpha + \beta \ln^2 (\delta)$ and $b = - 2 \beta \ln (\delta)$,
providing the explicit correlation among the high-energy parameters.
The above expansion, however, is not possible in the case of $\gamma$ real ($\neq$ 2)
so that the correlations are somewhat hidden in the nonlinear data reductions.

\item{4.} 
As commented before, the $\sigma_{HE}(s)$ component accounts for the rise
of the total cross-section at high energies and is associated with the Pomeron exchange.
In the standard soft Pomeron concept this term is expected
to increase with the energy, as is the case of the simple pole parametrization,
$s^{\epsilon}$ with $\epsilon$ slightly greater than zero \cite{pred,land,dl-a,dl-b}.

\item{5.} 
Another aspect of interest concerns the minimum value of the energy above which a given
parametrization is supposed to be applied, or valid. In this respect,
as we will discuss, the energy cutoff for data reduction, $\sqrt{s}_{min}$, plays a central role,
in connection with the corresponding energy scale,  $\sqrt{s_h}$.
Here, as in both COMPETE and PDG analyses, we have adopted  $\sqrt{s}_{min}$ = 5 GeV.

\end{description}

\vspace{0.3cm}

Based on the above characteristics, it is expected that, depending on the values
of $s_{min}$ and $s_h$, different physical interpretations can emerge in the
cases of $\gamma$ = 2 (fixed) and $\gamma$ real (not integer). 
Let us discuss these aspects and their connection with our fit results and
those from the COMPETE 2002 and PDG 2012 analyses.

\vspace{0.5cm}

\noindent
$\bullet$ $\gamma$ = 2 (fixed)

In this case, if $s_h < s_{min}$ then in the region of experimental data 
($\sqrt{s} \geq \sqrt{s}_{min}$), the high-energy component $\sigma_{HE}(s)$
increases with the energy, as expected in the standard concept of the soft Pomeron
contribution. That is the case of the 2012 PDG version since, from Table \ref{t1}:

\vspace{0.2cm}

\centerline{$s_h$ = 16.21 $\pm$ 0.16  GeV$^2$ $<$ $s_{min}$ = 25 GeV$^2$.}

\vspace{0.2cm}

On the other hand, if $s_h > s_{min}$ then in the interval 
$\sqrt{s}_{min} \leq \sqrt{s} \leq \sqrt{s_h}$, the  $\sigma_{HE}(s)$ component
decreases as the energy increases, suggesting a physical disagreement with the
standard soft Pomeron contribution. That, however, is the case of the 2002 COMPETE
result, since, from Table \ref{t1}:

\vspace{0.2cm}

\centerline{$s_h$ = 34.00 $\pm$ 0.54 GeV$^2$ $>$ $s_{min}$ = 25 GeV$^2$.}

\vspace{0.2cm}

The dependence of  $\sigma_{HE}(s)$ for the two cases above is shown
in Fig. \ref{f7} in the energy interval
4 GeV $\leq \sqrt{s} \leq$ 7 GeV that includes the energy cutoff $\sqrt{s}_{min}$
= 5 GeV.

In this respect, according to the COMPETE Collaboration \cite{compete1}:
``One must note that in some processes, the falling $\ln^2(s/s_0)$ term
from the triple pole at $s < s_0$ is important in restoring the degeneracy
of the lower trajectories at low energy. Hence the squared logarithm manifests itself not only
at very high energies, but also at energies below its zero."
However,
even accepting this argument on a decreasing Pomeron contribution in the physical
region considered, the fast 
rise of $\sigma_{HE}(s)$ as the energy
decreases below $\sqrt{s_h}$ (Figure \ref{f7}) and above the physical
threshold ($2m_p$), remains, in our opinion, unexplained.

A practical or even pragmatic consequence of this COMPETE result is  
the agreement between their 2002 prediction (with $\gamma$ = 2
in accordance with the Froissart-Martin bound) and the 7 TeV TOTEM measurement.
Or, in other words, this result
is directly connected to the rather large value of $s_h$. By contrast, the PDG result
with $\gamma$ = 2 and smaller $s_h$, which is consistent with a rise of
$\sigma_{HE}(s)$ in the whole interval of energy investigated,
 lies below the TOTEM datum (compare Figures
\ref{f1}  and \ref{f7}). 

Summarizing, with $\gamma$ = 2: 
(a) the COMPETE correctly describes the 7 TeV TOTEM datum,
but with a decreasing $\sigma_{HE}(s)$  contribution above the cutoff $s_{min}$  up to $s_h$ and a 
increasing contribution as the energy decreases below $s_{min}$
(strictly divergent as $s$ decreases); (b) in the PDG 2012 result, $\sigma_{HE}(s)$ 
increases in the whole energy-interval investigated (above the cutoff) but lies (within the uncertainty)
below the TOTEM datum.

\vspace{0.2cm}

\noindent
$\bullet$ $\gamma$ real (not integer)

For $\gamma$ not integer, as commented before, the $\sigma_{HE}(s)$ component
is not defined at $\sqrt{s} < \sqrt{s_h}$. The contribution starts at
$\sqrt{s} = \sqrt{s_h}$ with $\sigma_{HE}(s_h)$ = $\alpha$ and from this point on it increases
with the energy, as expected in the standard soft Pomeron concept.
With respect to our fit results, the $\sigma_{HE}(s)$ component is well defined
in the whole interval of energy investigated
since $\sqrt{s}_{min}$ = 5 GeV and in all data reductions
$s_h < s_{min}$ (Tables \ref{t2} and \ref{t3}).

\subsection{Conclusions on the best fit results}
\label{s43}

Based on the physical aspects related to the scale factor,
we now examine our unconstrained and constrained fit results
in connection with our representation (6). This discussion,
together with that in Subsec. \ref{s41},  will  lead us
to conclude that the best results are those obtained with the constrained variant.

With the unconstrained fits ($\delta$ or $s_h = \delta 4m_p^2$ free) we have obtained
for the $\sqrt{s}_{max}$ = 7 TeV ensemble $s_h \sim$ 13 GeV$^2$ (individual fit),
$s_h \sim$ 16 GeV$^2$ (global fit) and for the $\sqrt{s}_{max}$ = 8 TeV ensemble
$s_h \sim$ 0.6 GeV$^2$ (individual fit).
In these cases it seems difficult to devise a physical meaning
for the onset of the $\sigma_{HE}(s)$ component, because its value depends on 
the variant considered and data
analyzed. Moreover, as mentioned in Subsec. \ref{s32},
with the $\sqrt{s}_{max}$ = 8 TeV ensemble, the error matrix is not positive definite in the individual
fit and no convergence was obtained in the global fit.

On the other hand, in the case of constrained fits ($s_h = 4m_p^2$ fixed)
the $\sigma_{HE}(s)$ component starts at this threshold with 
$\sigma_{HE}(s_h) = \alpha$ and from this point on it increases with the energy,
namely $\sigma_{HE}(s) \geq \alpha$ at $\sqrt{s} \geq 2m_p$.
This threshold, $\sqrt{s_h} \sim 2$ GeV, is also below the energy cutoff, 
$\sqrt{s}_{min}$ = 5 GeV.
This situation seems physically meaningful to us in both phenomenological
and theoretical contexts. Furthermore, in all cases investigated with the constraint condition,
especially with the $\sqrt{s}_{max}$ = 8 TeV ensemble,
we have obtained full convergence in the data reductions and consistent statistical
results. As shown in Figure \ref{f6}, the values of $\gamma$ are also consistent (stable)
and lie around 2.2 - 2.3, in all cases investigated with the constrained variant.

This discussion and the points raised in the previous sections
favor, therefore, the results obtained with the constrained fits.
Among them, we select as our best results the individual and global fits
with the $\sqrt{s}_{max}$ = 8 TeV ensemble. Both indicate
a rise of the total cross-section that is faster than the log-squared behavior.

\section{Fits to Elastic Cross-Section Data}
\label{s5}

The total cross-section is related to the elastic cross-section in the forward direction
via the optical theorem. That has led us \cite{fms2} to explore the possibility of extending the
same analytical parametrization of the total cross-section, Eqs. (1)-(3), to the elastic (integrated)
cross-section data, $\sigma_{el}$. Based on unitarity, the same value of the exponent $\gamma$  obtained
for $\sigma_{tot}(s)$ is assumed for $\sigma_{el}(s)$. For a detailed discussion on this
assumption see section 3 in Ref. \citen{fms2}.

\subsection{Fit and results}
\label{s51}

The experimental data on $pp$ and $\bar{p}p$ scattering below 7 TeV have been extracted from the PDG 
database
\cite{pdg12}, without any kind of data selection or sieve procedure. The dataset includes
also the recent TOTEM results (four points at 7 TeV (Ref. \citen{totem5}) and one point 
at 8 TeV (Ref. \citen{totem6})). 
Statistical and systematic errors have been added in quadrature.

As feedback for initializing the parametrization of $\sigma_{el}$ data
we consider here the values of the parameters from our selected fit results
to $\sigma_{tot}$ with both  $\gamma$  and $s_h$ fixed (constrained fits in Table \ref{t3}). 
We notice that the data reductions using as initial values the results from either the 
individual or global fits
are similar. In what follows we focus mainly on the global case. The results with
$\gamma$ = 2.23 and $s_h$ = 3.521 GeV$^2$ (fixed) are displayed in Table \ref{t7}
and in Figure \ref{f8} (up) together with the evaluated uncertainty region.

\begin{table}[ht]
\tbl{Fit results to the elastic cross-section data with feedbacks
from the global constrained fit result to $\sigma_{tot}$ data in the
case of the $\sqrt{s}_{max}$ = 8 TeV ensemble (Table \ref{t3}, last column).
Units as in Table \ref{t1}.}
{\begin{tabular}{cc}
\hline
$a_1$        & \qquad \quad      32.6 $\pm$ 4.7    \\
$b_1$        & \qquad \quad     0.579 $\pm$ 0.044   \\
$a_2$        & \qquad \quad       0.9 $\pm$ 1.1   \\
$b_2$        & \qquad \quad      0.41 $\pm$ 0.31   \\
$\alpha$     & \qquad \quad     4.74 $\pm$ 0.15   \\
$\beta$      & \qquad \quad    0.0380 $\pm$ 0.0010 \\
$\gamma$     & \qquad \quad         2.23 (fixed)       \\
$s_h$        & \qquad \quad        3.521 (fixed)     \\
\hline
DOF          & \qquad \quad            102          \\
$\chi^2$/DOF & \qquad \quad            1.55        \\
$P(\chi^2)$  & \qquad \quad 3.33$\times$10$^{-4}$ \\
\hline
Figure:      & \qquad \quad        8              \\
\hline
\end{tabular}
\label{t7}}
\end{table}

From Table  \ref{t7}, the value of the $a_2$ parameter is statistically consistent with zero.
This is a consequence of the equality of the $pp$ and $\bar{p}p$ elastic cross-sections
data at low energies. We have checked that letting $a_2$ = 0, the same fit result is
obtained. However, we notice that the statistical quality of the fit is not so good: large reduced
$\chi^2$ and small integrated probability. Moreover, from Figure \ref{f8}, the fit uncertainty
region barely reaches the extremum of the lower error bar of the TOTEM result at 8 TeV.
On statistical grounds, since this point constitutes a high-precision measurement (defining the experimental
information at the highest energy), the somewhat 
low fit quality may be associated with the underestimation of this datum by the fit result.
We shall return to this important point related to the 8 TeV TOTEM data in section
\ref{s61}.

Despite the limitations on the statistical quality of the fit, from Figure \ref{f8} (up),
the global description of the experimental data seems satisfactory, including,
within the uncertainties, the lower error bars of the four TOTEM results at 7 TeV.
If we accept this data reduction as a reasonable description of the experimental data,
the results, together with that obtained for the total cross-section, allow us
to predict the ratio between the elastic and total cross-section as function of the
energy. The result, within the uncertainties, is shown in Figure \ref{f8} (down), together 
with the experimental
data. Using the $s$-channel unitarity, we have also included in this figure the result 
from the estimations of the total cross-section and the inelastic cross-section 
($\sigma_{inel}$) at 57 TeV by the Auger Collaboration \cite{auger}.

\subsection{Asymptotic ratios}
\label{s52}

The asymptotic ratio between the elastic and total cross-sections can be 
evaluated from parametrization (1)-(3). Denoting the parameters $\beta$ associated with the $\sigma_{tot}$ and $\sigma_{el}$
fits by the corresponding indexes, for $s \rightarrow \infty$, we have
\begin{eqnarray} 
\frac{\sigma_{el}}{\sigma_{tot}}  \rightarrow \frac{\beta_{el}}{\beta_{tot}}.
\nonumber
\end{eqnarray} 
From Tables \ref{t3} and \ref{t7} and the $s$-channel unitarity, we obtain
\begin{eqnarray} 
\frac{\sigma_{el}}{\sigma_{tot}}  \rightarrow  0.31 \pm 0.13
\qquad
\mathrm{and}
\qquad
\frac{\sigma_{inel}}{\sigma_{tot}}  \rightarrow  0.69 \pm 0.13,
\nonumber
\end{eqnarray} 
a result which is not in agreement with the naive black-disk model (limit 1/2), but
statistically consistent, within the uncertainties, with rational limits
\begin{eqnarray} 
\frac{\sigma_{el}}{\sigma_{tot}}  \rightarrow  \frac{1}{3}
\qquad
\mathrm{and}
\qquad
\frac{\sigma_{inel}}{\sigma_{tot}}  \rightarrow  \frac{2}{3},
\end{eqnarray}
as already obtained in our previous analysis \cite{fms2}, where  only the first TOTEM measurement
at 7 TeV has been included. We note that from the individual fit to $\sigma_{tot}$
($\gamma$ = 2.30 and $s_h$ = 3.521 GeV$^2$ fixed) we obtain $\sigma_{el}/\sigma_{tot}$
$\rightarrow$ 0.301 $\pm$ 0.098, a result also in agreement with (8), within the
uncertainties.

These rational limits contrast with the prediction from the model-dependent 
amplitude analysis
by Block and Halzen, which indicates the black-disk limit for both ratios \cite{bhbd}. 
However, the rational limits are not in disagreement
with a saturation of the Pumplim bound \cite{pump,suk},
\begin{eqnarray} 
\frac{\sigma_{el}}{\sigma_{tot}}  + \frac{\sigma_{diff}}{\sigma_{tot}}
\leq \frac{1}{2},
\nonumber
\end{eqnarray} 
where $\sigma_{diff}$ is the cross-section associated with the soft diffractive
processes (single and double dissociation). This saturation and the
rational limits corroborate the
recent phenomenological arguments by Grau \textit{et al}. who
attribute the black-disk limit to the combination of the elastic
and diffractive processes, calling also the attention to the possibility
of the limiting value 1/3 \cite{grau}.
If that is the case, our results predict
\begin{eqnarray} 
\frac{\sigma_{diff}}{\sigma_{tot}}  \rightarrow \frac{1}{6}
\qquad
\mathrm{as}
\qquad
s \rightarrow \infty.
\nonumber
\end{eqnarray}

\section{On a Fast Rise of the Total Cross-Section}
\label{s6}

Our results with both ensembles ($\sqrt{s}_{max}$ = 7 TeV and 8 TeV) indicate the
possibility of a rise of
$\sigma_{tot}$ faster than the log-squared behavior. In addition, there seems to be
some interesting aspects related to the data at the highest LHC energy that deserve further comments.
In this section we first discuss
some features of the 8 TeV TOTEM data ($\sigma_{tot}$ and $\sigma_{el}$), as compared with
those at 7 TeV and in the region below this energy. After that, we present a few conjectures
related to the possibility of a fast increase of the total cross-section at the LHC energy region and beyond.

\subsection{The 8 TeV TOTEM data}
\label{s61}

From Tables \ref{t2} and \ref{t3}, in going from the $\sqrt{s}_{max}$ = 7 TeV ensemble
to the $\sqrt{s}_{max}$ = 8 TeV ensemble, we can note a slight increase in the value
of the $\gamma$ parameter (although consistent within the uncertainties
in the constrained case). That may suggest a rise of $\sigma_{tot}$ in the
7 - 8 TeV region faster than that observed up to 7 TeV. In this respect we draw the
attention to the results that follows.

\begin{figure}
\centering
\epsfig{file=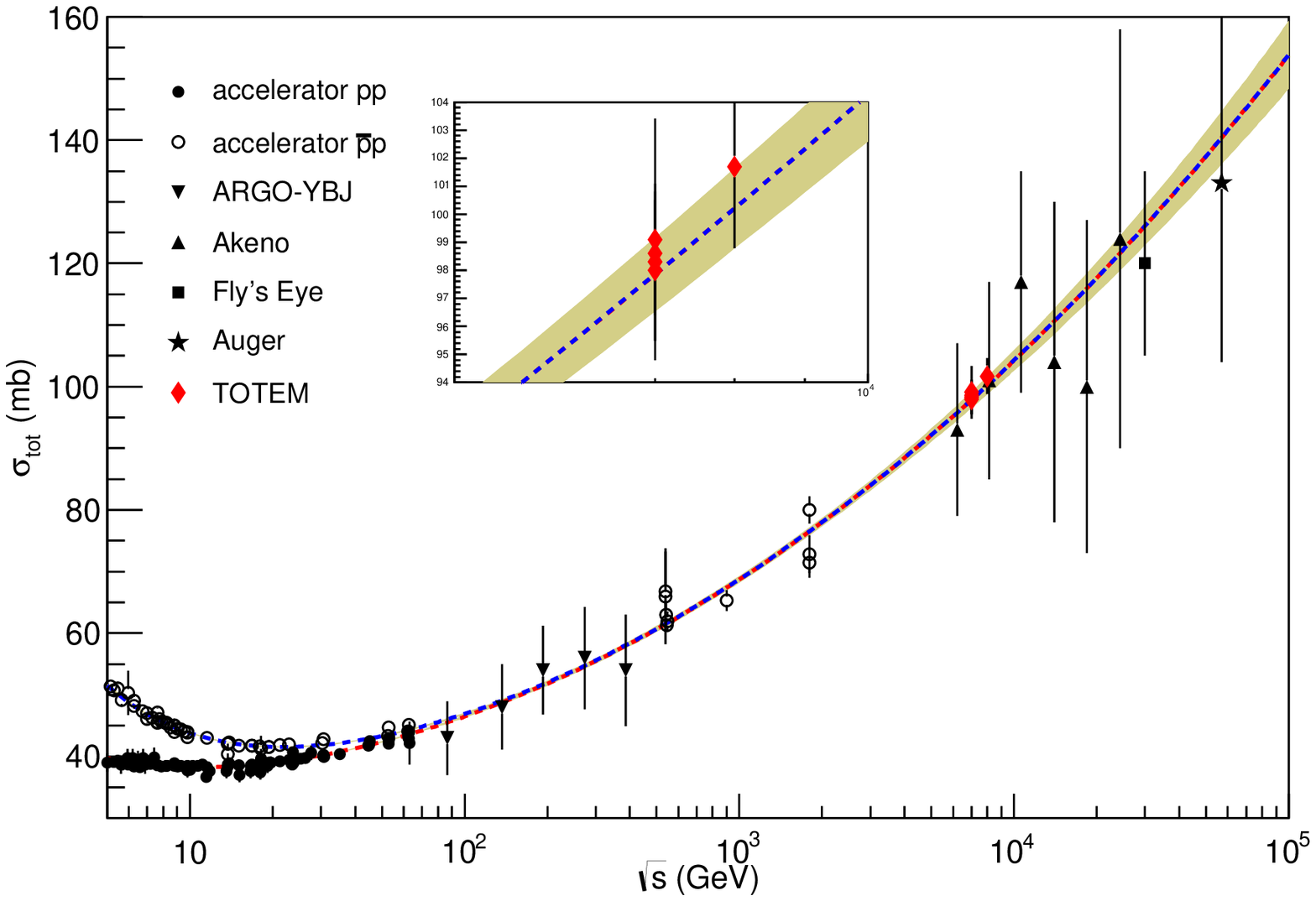,width=15cm,height=8.5cm}
\epsfig{file=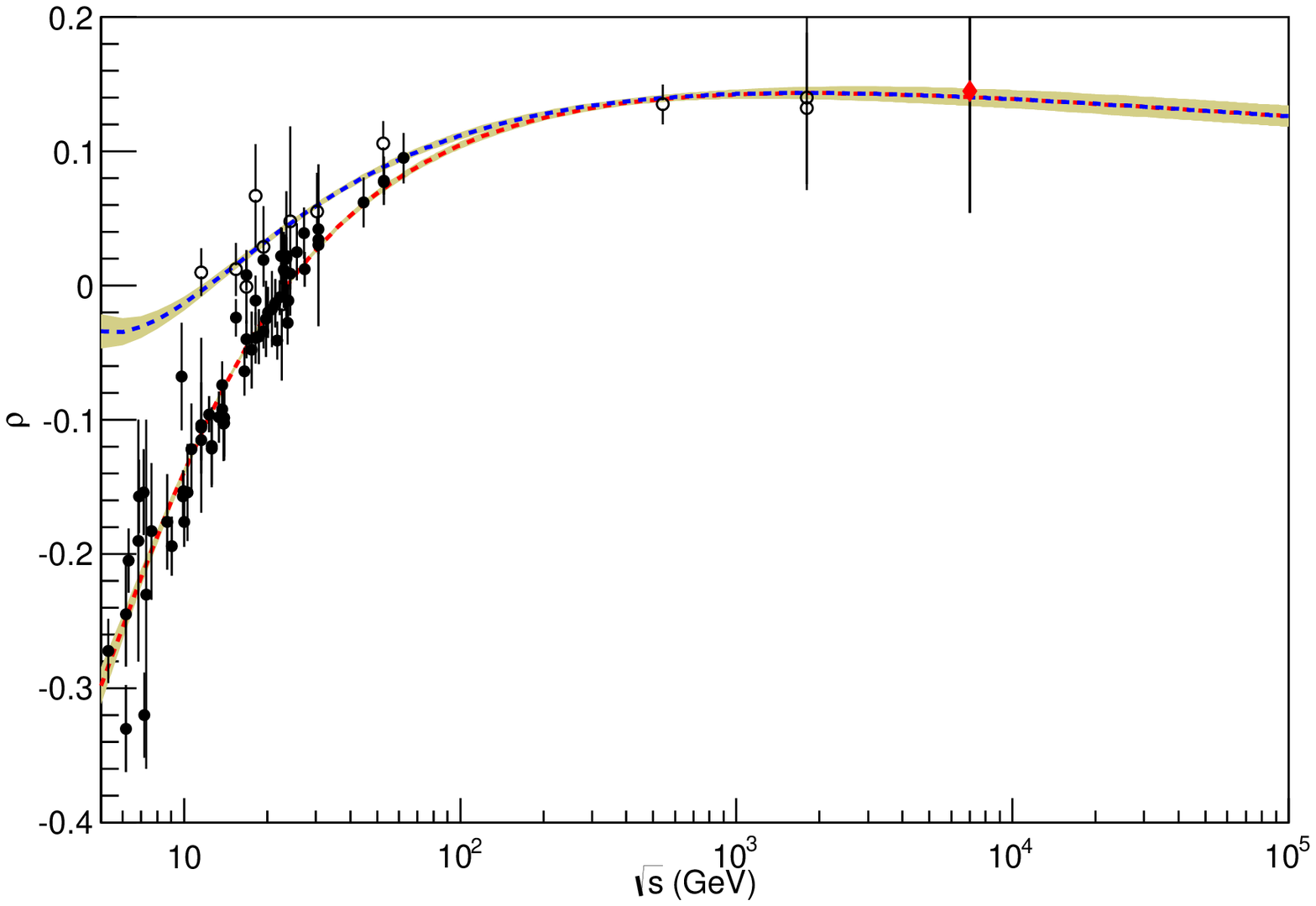,width=15cm,height=8.5cm}
\caption{Global constrained fit results with the $\sqrt{s}_{max}$ = 8 TeV
ensemble (fifth column in Table \ref{t3}).}
\label{f4}
\end{figure}

\begin{figure}
\centering
\epsfig{file=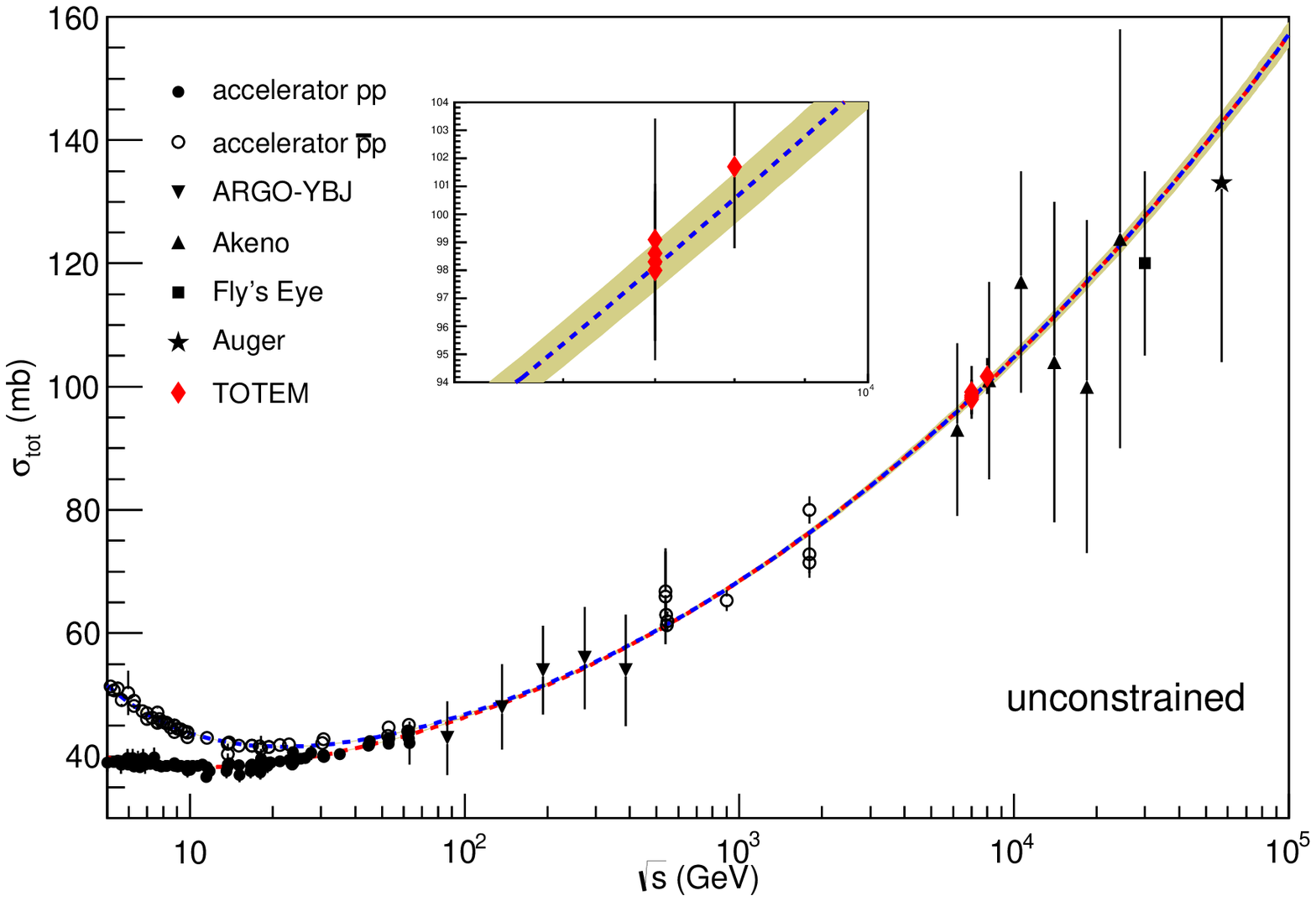,width=15cm,height=8.5cm}
\epsfig{file=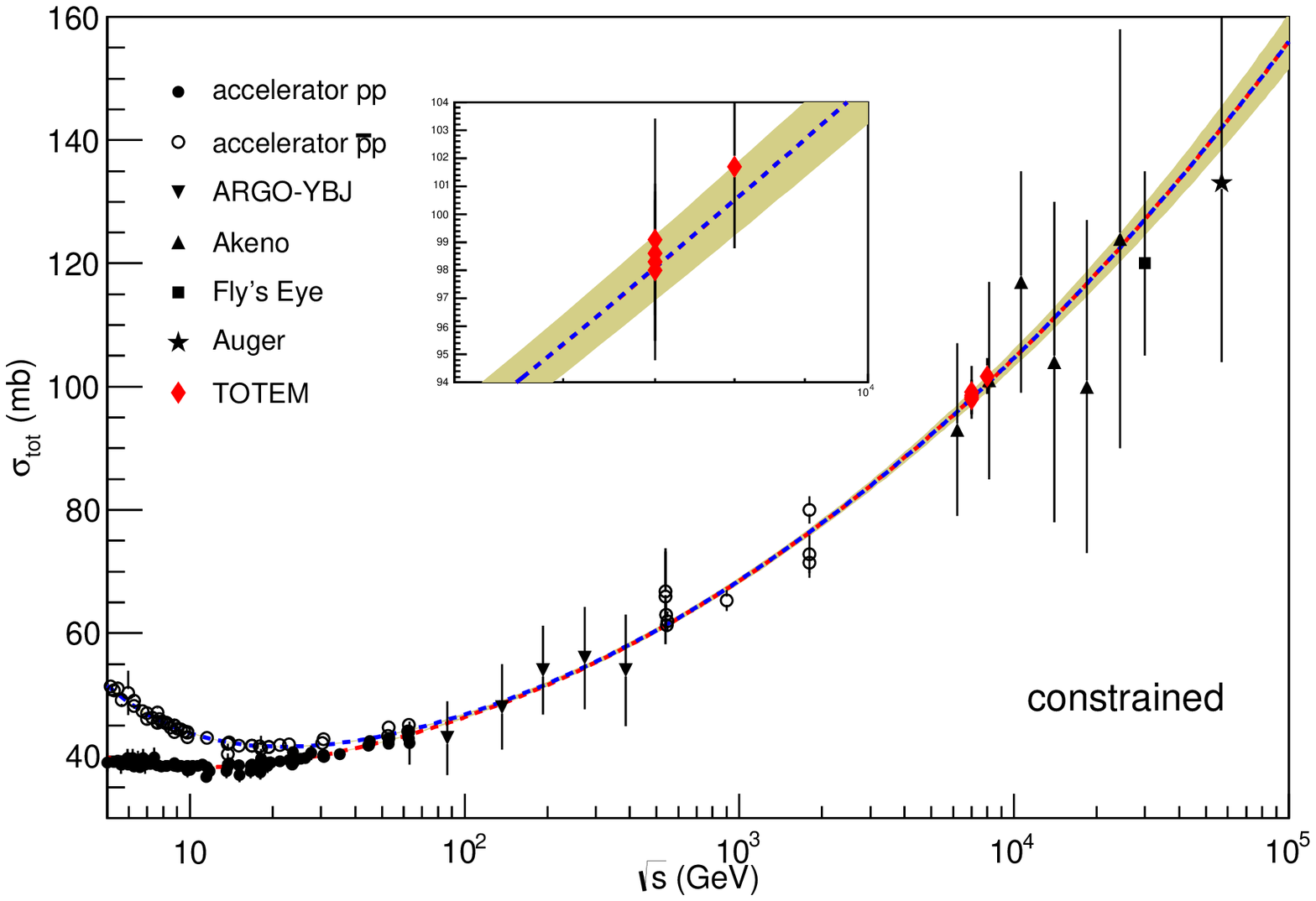,width=15cm,height=8.5cm}
\caption{Individual fit results to $\sigma_{tot}$ data with the $\sqrt{s}_{max}$ = 8 TeV 
ensemble and unconstrained (up) and constrained (down) data reductions
(second column and fourth column in Table \ref{t3},
respectively).}
\label{f5}
\end{figure}

\begin{figure}
\centering
\epsfig{file=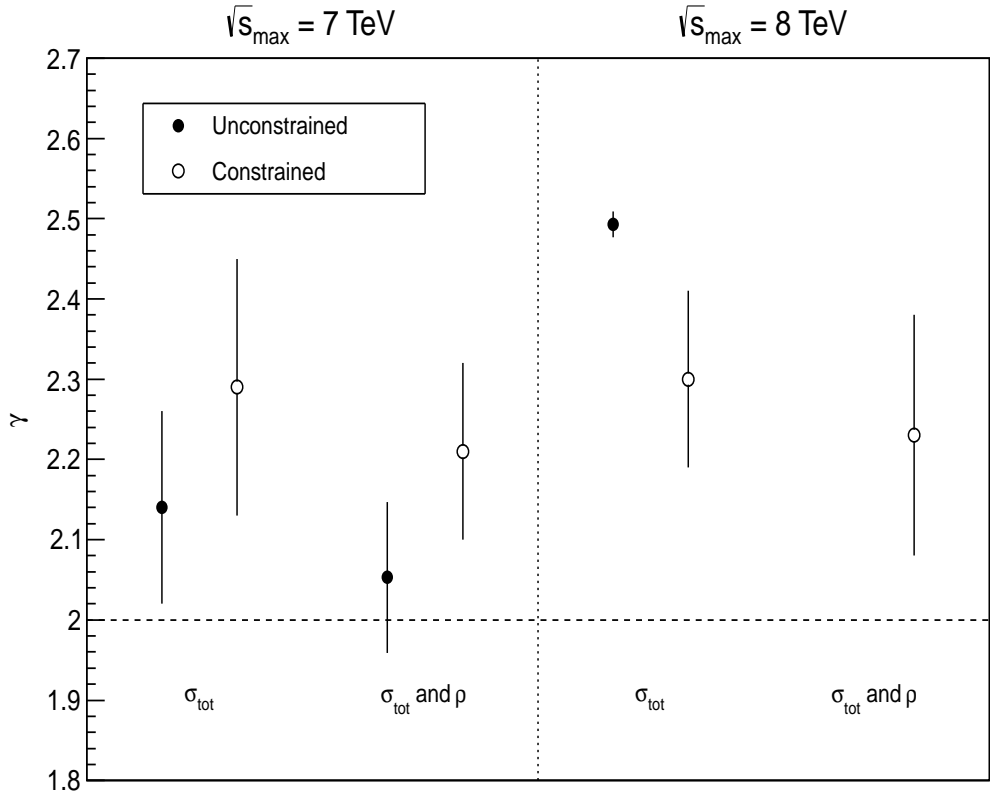,width=15cm,height=12cm}
\caption{Results obtained here for the exponent $\gamma$ as a free parameter
in different data reductions through parametrization (1)-(3)
and (1)-(5):
ensembles $\sqrt{s}_{max}$ = 7 TeV (Table \ref{t2}) and $\sqrt{s}_{max}$ = 8 TeV (Table \ref{t3}), constrained and unconstrained fits,
individual ($\sigma_{tot}$) and global ($\sigma_{tot}$ and $\rho$) fits.}
\label{f6}
\end{figure}

\begin{figure}
\centering
\epsfig{file=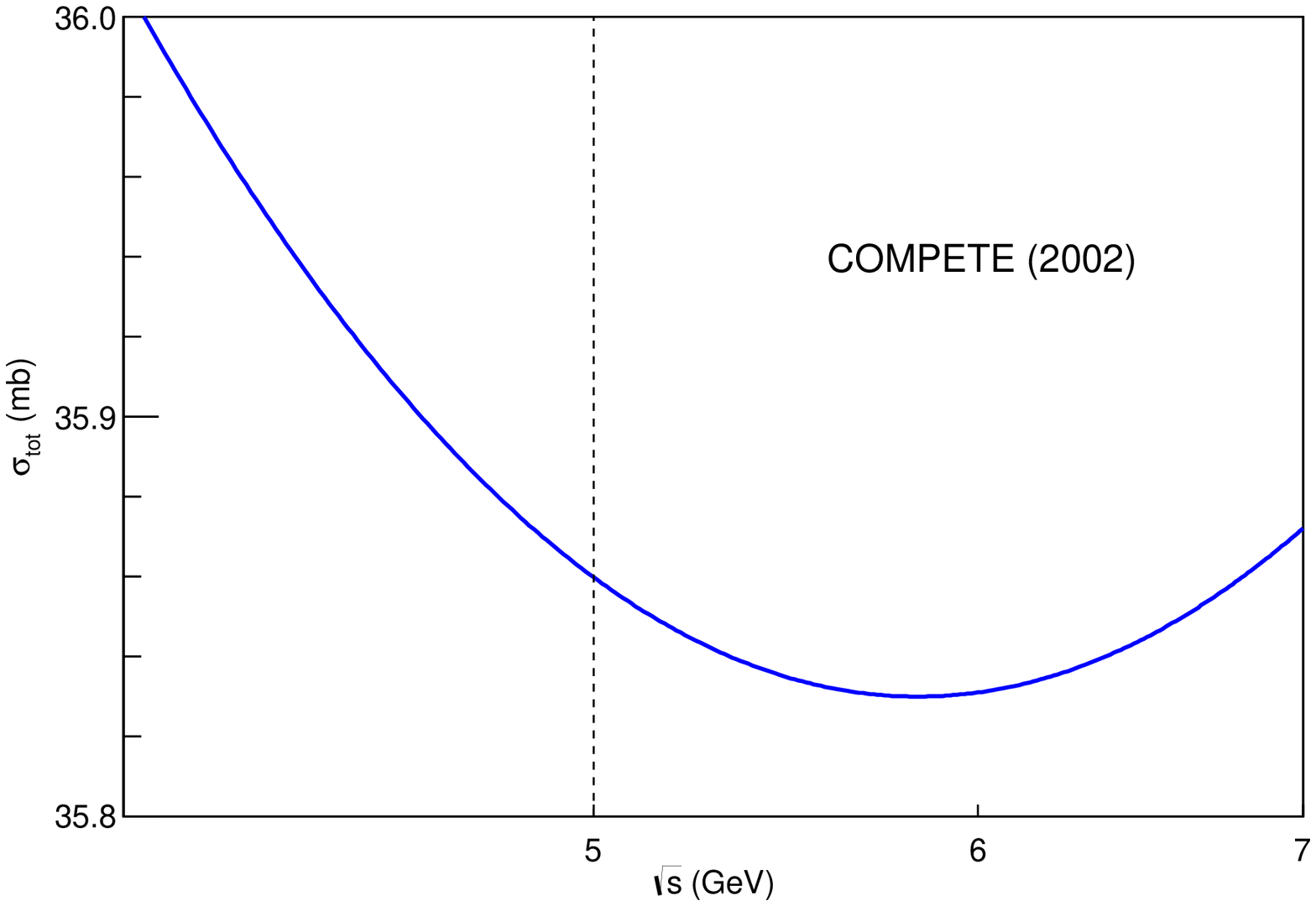,width=15cm,height=8.5cm}
\epsfig{file=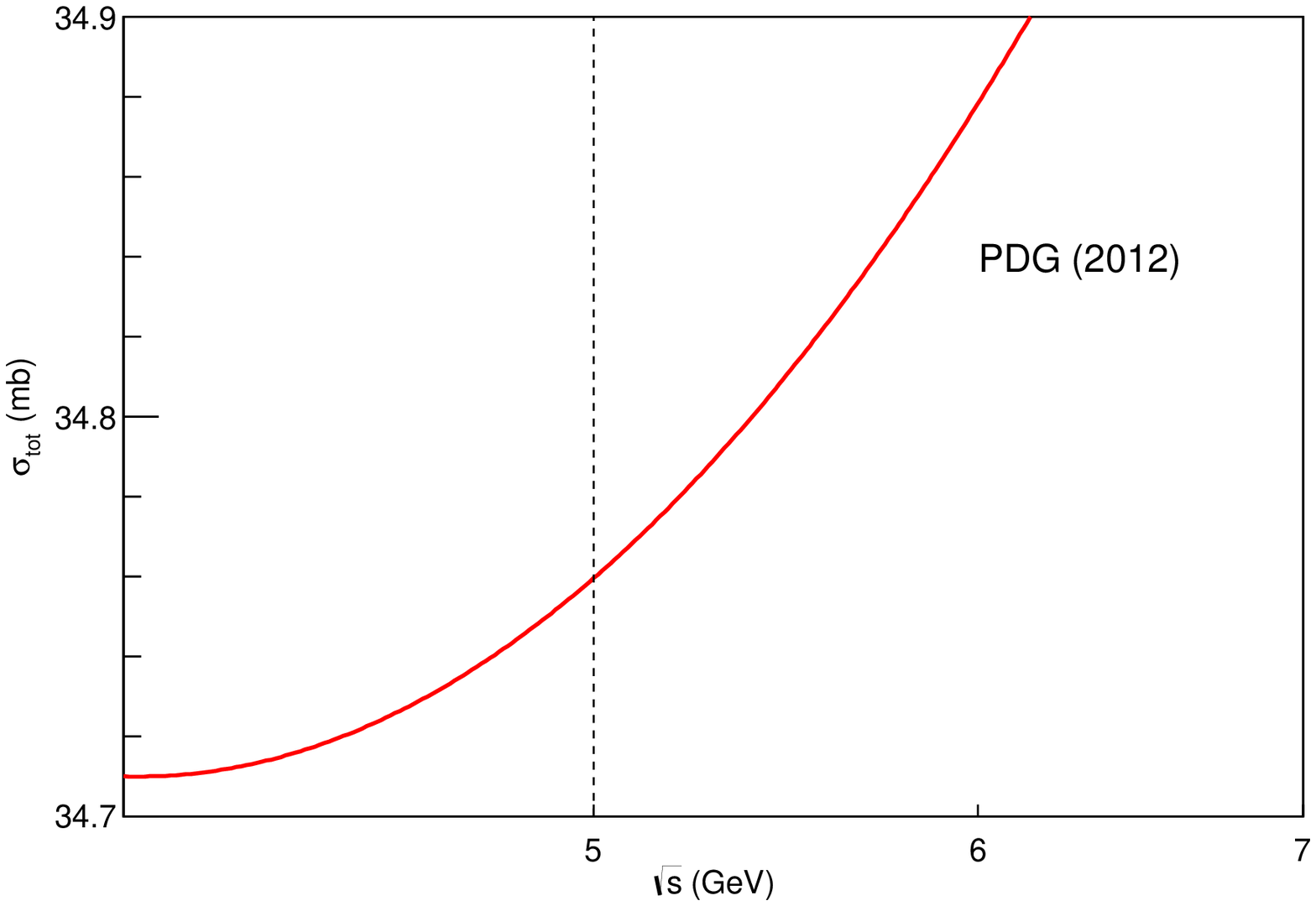,width=15cm,height=8.5cm}
\caption{High energy contribution $\sigma_{HE}(s) =
\alpha\, + \beta\, \ln^2 (s/s_h)$ from the COMPETE 2002 and PDG 2012
analyses, around the energy cutoff $\sqrt{s}_{min}$ = 5 GeV
(parameters from Table \ref{t1}).}
\label{f7}
\end{figure}

\begin{figure}
\centering
\epsfig{file=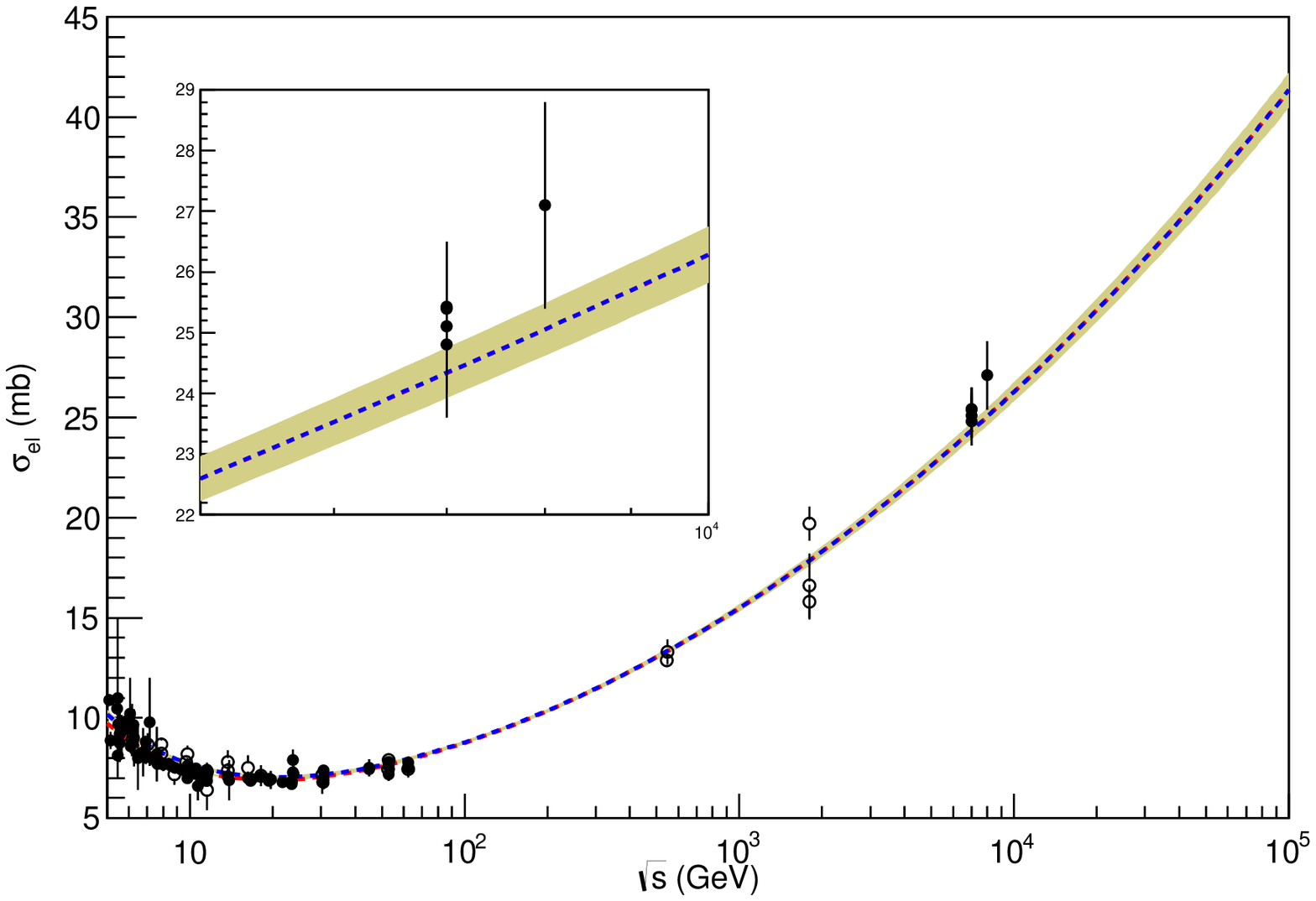,width=15cm,height=8.5cm}
\epsfig{file=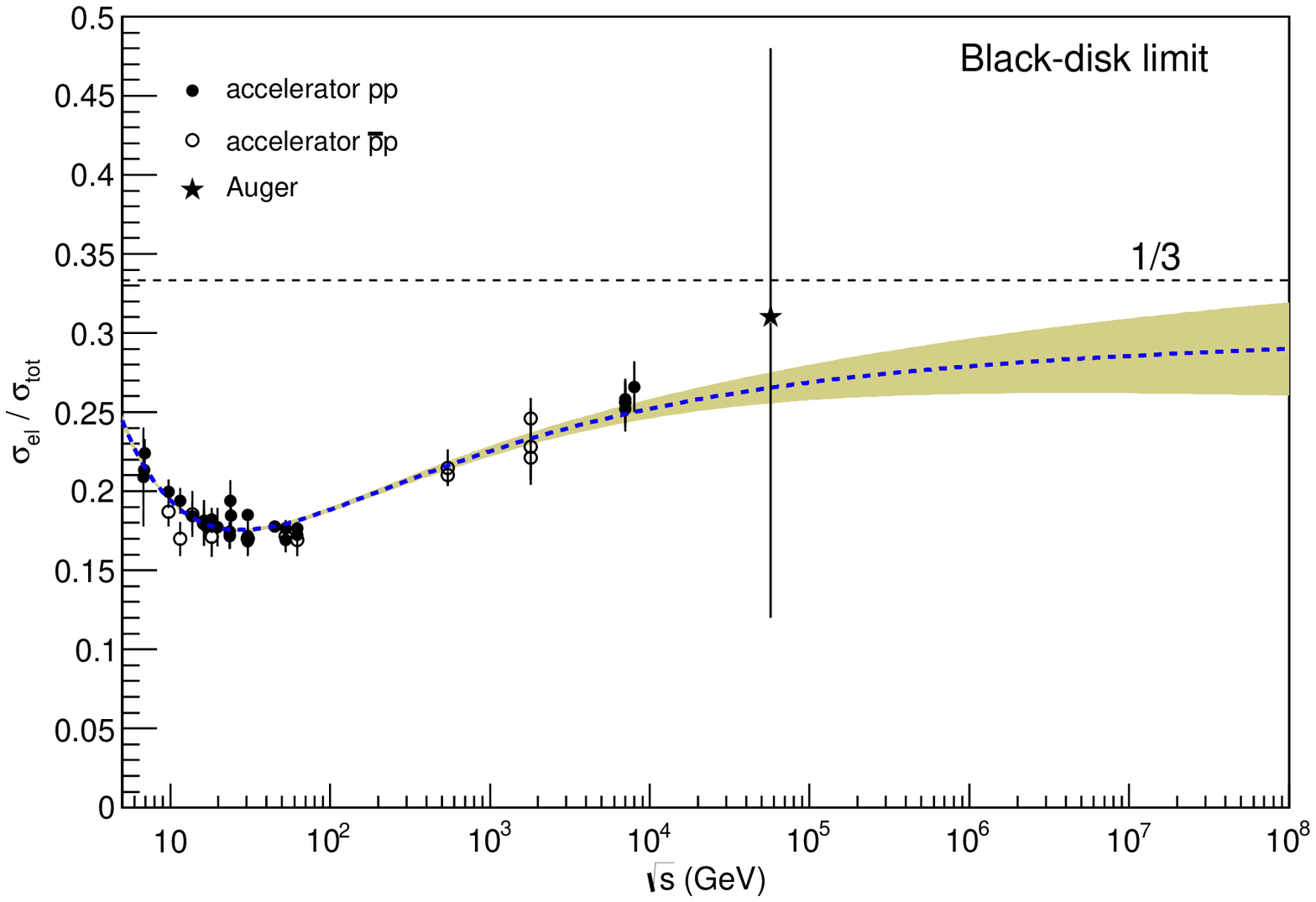,width=15cm,height=8.5cm}
\caption{Fit result for the elastic cross-section data (up)
and predictions for the ratio between the elastic and total cross-sections
(down).}
\label{f8}
\end{figure}

\begin{description}

\item{1.}
In all data reductions for $\sigma_{tot}$ presented here, the central values of the TOTEM data
at 7 TeV are well described within the uncertainties (Figures \ref{f2} - \ref{f5}).

\item{2.}
In the case of the 8 TeV datum, the same results present good agreement only with 
the lower error bar and, in general, the fit uncertainty does not reach the central value.
With our selected constrained results the uncertainty region includes the central
value in the case of the individual fit (Figure \ref{f5}) and only barely reaches this point in
the case of the global fit (Figure \ref{f4}).

\item{3.}
Analogous effects can be observed in the case of the $\sigma_{el}$ data:
within the uncertainties the fit results are consistent with the lower error
bars at 7 TeV, but reaches only the extremum of the lower bar at 8 TeV.

\end{description}

Therefore, even with $\gamma$ greater than 2 and the scaling parameter fixed or free,
the fit results are not in statistical agreement with the 8 TeV TOTEM data
on $\sigma_{tot}$ and $\sigma_{el}$:
the fits somewhat underestimate the high-precision experimental values,
suggesting a rise faster than expected.
In this respect, some quantitative inferences
may be instructive, even if only in a limited context,
as discussed in what follows.

First, although associated with different variants, the results
here obtained for the $\gamma$ parameter ($\gamma_i \pm \Delta \gamma_i$)
can be used to provide a quantitative information on typical values
associated, separately, with the ensembles $\sqrt{s}_{max}$ = 7 TeV 
(four points) and $\sqrt{s}_{max}$ = 8 TeV (three points). We
have considered two evaluations, either a weighted mean (with
weights $1/[\Delta \gamma_i]^ 2$) or a fit by
a constant function (MINUIT). The results are displayed in 
Table \ref{t8} showing that, from ensemble $\sqrt{s}_{max}$ = 7 TeV to 
$\sqrt{s}_{max}$ = 8 TeV, both evaluations indicate an 
increase in the value of $\gamma$ around 16 \%.

\begin{table}[ht]
\tbl{Global estimates for the average values of the parameter $\gamma$ from fits with ensembles
$\sqrt{s}_{max}$ =  7 TeV (Table \ref{t2}, four points) and $\sqrt{s}_{max}$ = 8 TeV
(Table \ref{t3}, three points).}
{\begin{tabular}{c c c}
\hline
Ensemble    \quad           &  \quad   Weighted mean      & \quad Constant function fit     \\
\hline
 $\sqrt{s}_{max}$ = 7 TeV   & 2.146  $\pm$ 0.003   & 2.15  $\pm$ 0.06      \\
 $\sqrt{s}_{max}$ = 8 TeV  & 2.4861 $\pm$ 0.0003  & 2.49  $\pm$ 0.02  \\
\hline
\end{tabular}
\label{t8}}
\end{table}

\begin{figure}
\centering
\epsfig{file=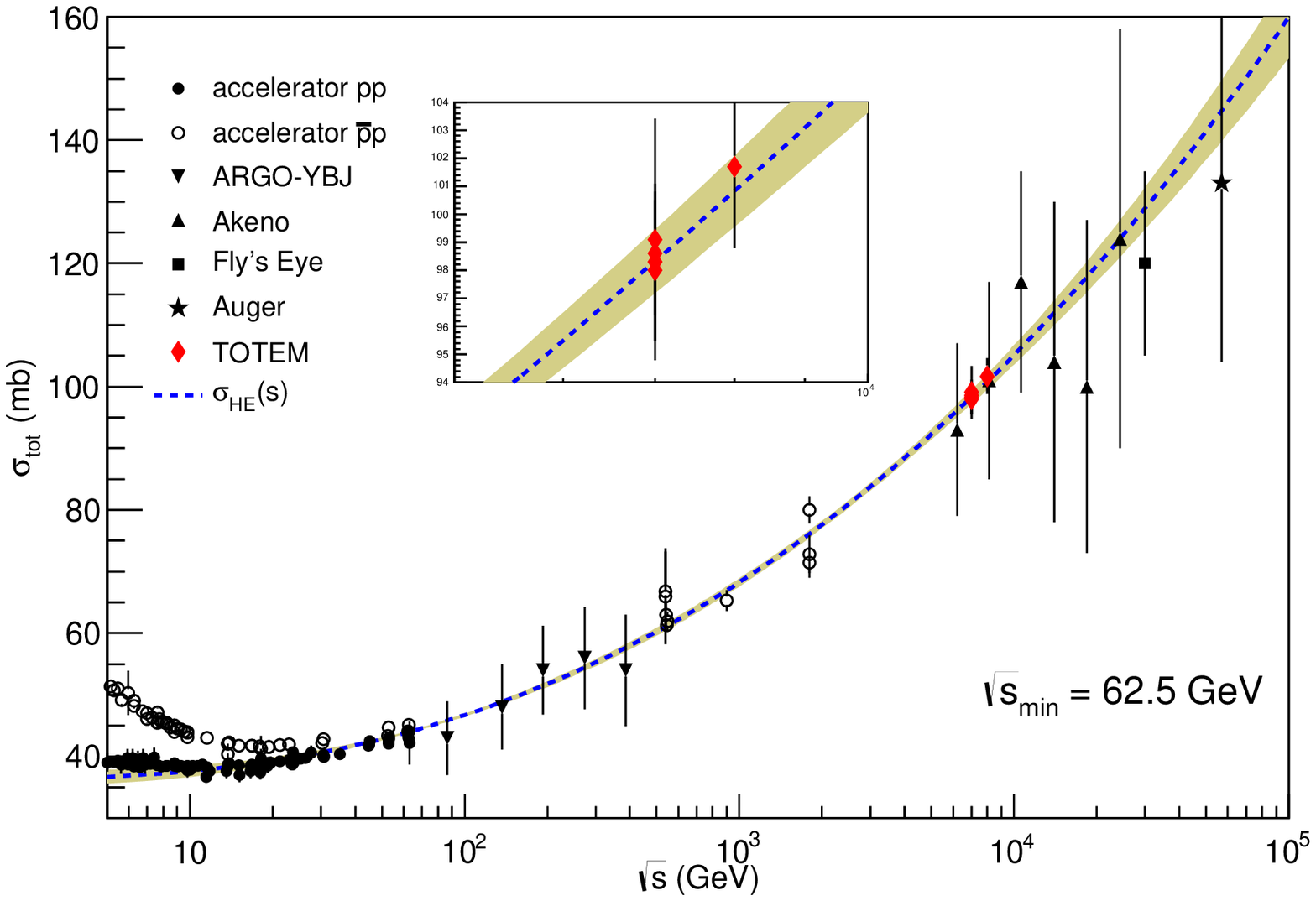,width=15cm,height=8.5cm}
\epsfig{file=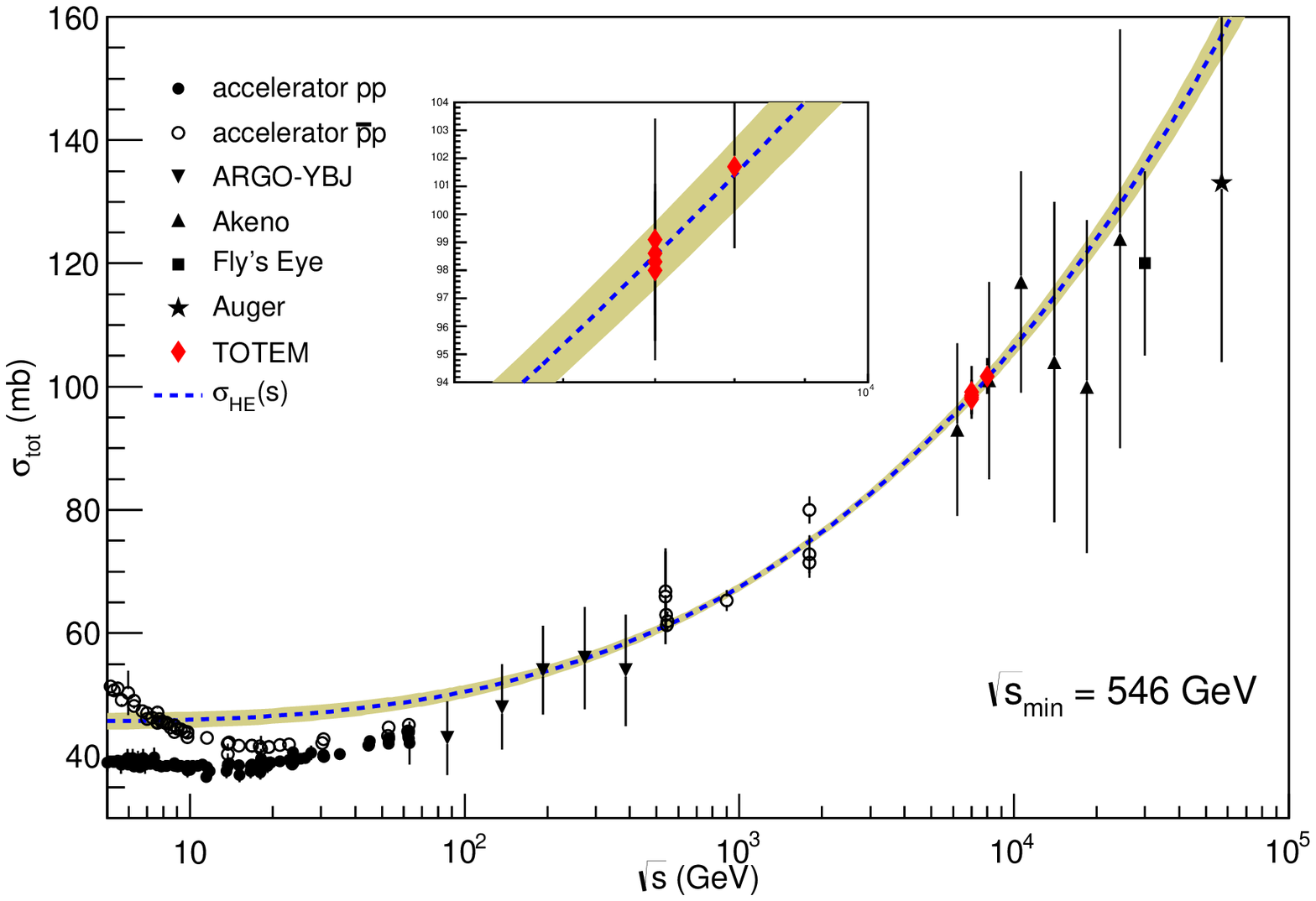,width=15cm,height=8.5cm}
\caption{Results of the constrained fit ($s_h$ fixed) to ensemble $\sqrt{s}_{max}$ = 8 TeV
through parametrization $\sigma_{\mathrm{HE}}(s)$, Eq. (3), and two energy cutoffs, $\sqrt{s}_{min}$ = 62.5 GeV and $\sqrt{s}_{min}$ = 546 GeV (Table 9).}
\label{f9}
\end{figure}

Second, in order to get some quantitative information directly
related to the highest energy region, we have developed fits
to $\sigma_{tot}$ data with only the high-energy parametrization
$\sigma_{HE}$, Eq. (3), but applied to datasets with different energy cutoffs: 
$\sqrt{s}_{min}$ = 62.5 GeV (CERN-ISR), 546 GeV (CERN-Collider) and 
1.8 TeV (Fermilab).
As already selected, we have considered the constrained variant
($s_h = 4m_p^2$ fixed) with $\sqrt{s}_{max}$ = 8 TeV. In this case
the parametrization has only three free fit parameters, 
namely $\alpha$, $\beta$ and $\gamma$. 
For the first cutoff ($\sqrt{s}_{min}$ = 62.5 GeV) we have used as feedback 
the values of the parameters obtained in the individual fit to $\sigma_{tot}$
data in the constrained case and ensemble $\sqrt{s}_{max}$ = 8 TeV
(fourth column in Table 3). Then, the fit result has been used as feedback
for the second cutoff ($\sqrt{s}_{min}$ = 546 GeV) and the same process for the
third one. The results with the first two cutoffs are displayed in Table \ref{t9} and Figure 
\ref{f9} (for $\sqrt{s}_{min}$ = 1.8 TeV the results are similar to those obtained with
$\sqrt{s}_{min}$ = 546 GeV).

From Figure \ref{f9} we note that in the case of $\sqrt{s}_{min}$ = 62.5 GeV
the fit result with only three parameters is in plenty agreement with all the $pp$ and $\bar{p}p$
experimental data above $\sim$ 30 GeV, describing also the $pp$ data at
lower energies. The TOTEM data at 7 and 8 TeV are also described within the uncertainties
and in this case $\gamma \approx$ 2.5 (Table \ref{t9}). With the cutoff $\sqrt{s}_{min}$ = 546 GeV,
the fit is in agreement with the high-energy data (above $\sim$ 100 GeV) and the
TOTEM data is quite well described (especially at 8 TeV); however, in this case
$\gamma \approx$ 3.3 (Table \ref{t9}).

We understand that all the aforementioned results and discussions seem to suggest 
a fast unexpected rise of the cross-sections from 7 to 8 TeV
as compared with the region below 7 TeV.

\begin{table}[ht]
\tbl{Constrained fits ($s_h$ = 3.521 GeV$^2$ fixed) to ensemble $\sqrt{s}_{max}$
= 8 TeV with parametrization $\sigma_{HE}(s)$, Eq. (3), and two different 
energy cutoffs ($\sqrt{s}_{min}$).}
{\begin{tabular}{c c c }\hline
$\sqrt{s}_{min}$:\qquad  & \qquad     62.5 GeV      & \qquad      546 GeV          \\
\hline
 $\alpha$ (mb) & 36.40 $\pm$ 1.24   & 45.68  $\pm$ 1.34      \\
 $\beta$  (mb) & 0.062 $\pm$ 0.031  & 0.0054 $\pm$ 0.0014  \\
 $\gamma$ & 2.46  $\pm$ 0.17   & 3.282  $\pm$ 0.088   \\
\hline
 DOF      &        17          &         8            \\
 $\chi^2/$ DOF &   1.70        &        1.04          \\
 $P(\chi^2)$   &   0.035       &       0.401          \\ 
\hline
Figure:        &    9          &        9         \\
\hline
\end{tabular}
\label{t9}}
\end{table}

\subsection{Some conjectures}
\label{s62}

At this point, we could conjecture (if not speculate) on the implication
of a possible increase of $\sigma_{tot}$ faster than $\ln^2{s}$.
One possibility points to a power-like
behavior $s^{\epsilon}$, $\epsilon >$ 0, which 
has always been an important and representative approach
\cite{dl0,dl1,dl2,dl3}.
Predictions from unitarized models, developed nearly 10 years ago and consistent with the  
first 7 TeV TOTEM datum, are discussed, for example, in Refs. \citen{kope1,kope2,petro1a,petro1b,petro2}.

A faster-than-squared-logarithm rise could also indicate the onset of some new physics effect
at the LHC energy region. For example, one possible explanation for the short
penetration depth, recently observed in ultra-high-energy cosmic rays (UHECRs)
around 100 TeV, is just an increase of the proton cross-section faster than
the extrapolations from models, which have been tested only at lower energies 
(see, for example, Refs. \citen{piran} and \citen{deus} and references therein).
These conjectures are not in disagreement with the recent theoretical
arguments by Azimov  \cite{azimov1,azimov2,azimov3}.

At last, in contrast to an effective violation of the Froissart-Martin bound,
a fast rise of the total cross-section may also be associated
with some local effect at the LHC energy region and/or beyond, so that, asymptotically, 
the  bound might remain valid. In that case, $\gamma$ could represent a kind
of effective exponent, depending on the energy and, possibly, associated
with sums of different high-energy contributions.
However, if constituting only a local effect our asymptotic results for the ratios
involving the cross-sections might not be valid.

\section{Conclusions and Final Remarks}
\label{s7}

In 2002, Barone and Predazzi stated (Ref. \citen{pred}, page 140):

\begin{quote}
``The issue of the \textit{exact} growth with energy of the total cross-sections
is both delicate and unresolved; the mild growth of total cross-sections could
be simulated by essentially any form and logarithmic physics is exceedingly 
difficult to resolve in a clear cut way."
\end{quote}

The aim of this paper has been to take one more step in our investigation on the
rise of the total hadronic cross-section at high energies. As in our previous
analyses \cite{fms1,fms2} we have employed the analytical parametrization
introduced by Amaldi \textit{et al}., with the exponent $\gamma$
in the high-energy leading logarithm contribution treated as a free real parameter
in nonlinear data reductions.

Here, the main points consisted in an updated analysis (including in the dataset
the recent high-precision TOTEM measurements at 7 TeV and 8 TeV) and a discussion
on the correlation, practical role and physical meaning associated to the
exponent $\gamma$ and the energy scale factor $s_h$. As in our previous works, we have 
considered different variants, involving two ensembles, individual/global fits
and unconstrained/constrained fits. As feedbacks for the nonlinear data reductions
we have used the ``conservative" values obtained by the PDG in the recent 2012 
Review of Particle Physics edition.

In section \ref{s4} we have discussed all the fit results, indicating the advantages
of the constrained fits ($s_h = 4m_p^2$ fixed) in both phenomenological and theoretical
contexts and noticing also the statistical consistence of the fit results. In particular,
we have concluded that the \textit{constrained fits with the $\sqrt{s}_{max}$ = 8 TeV
ensemble} (individual and global cases) represent our best results
(Table \ref{t3}, fourth and fifth columns and Figures 4 and 5).
They indicate
a rise of the total hadronic cross-section faster than the log-squared bound at the
LHC energy region. 
The results and predictions for the $pp$ total cross-section at energies of
interest are displayed in the last columns of Tables \ref{t4} and \ref{t5}.
A critical discussion on the COMPETE 2002 prediction for the total cross-section
and the recent 2012 PDG result has been also presented.

With the selected results mentioned above, extensions of the parametrization to fit the 
elastic cross-section data,
with fixed $\gamma$, have led to almost satisfactory results. Asymptotic limits for the
ratios between elastic/total and inelastic/total cross-sections indicate
consistence with 1/3 and 2/3, respectively (in agreement with our previous
result \cite{fms2}).

We have called the attention to a possible fast rise of the  cross-section between
7 TeV and 8 TeV, according to the TOTEM results.
We have conjectured that a fast rise might be connected with
the onset of some new phenomena and have also speculated on the possible connection with
the short penetration depth recently observed in UHECRs.
If these effects have a local character (finite energies), there might be no contradiction
with the Froissart-Martin bound, since it has been derived for the asymptotic energy  limit,
$s \rightarrow \infty$.

At last, we mention our recently updated \textit{comparative} analysis \cite{ms2},
which includes also fits with either $\gamma$ = 2 (fixed) or a simple pole parametrization for the
high-energy contributions (namely $s^{\epsilon}$, $\epsilon >$ 0). 
Beyond further discussions on the effects associated with the parameters $\gamma$
and $s_h$ (not present in the simple pole parametrization), the results complement and
corroborate those presented here and in the previous works \cite{fms1,fms2}.

Our final conclusion, as we have stressed, is that the
rise of the total hadronic cross-section at the highest energies still constitutes an open problem,
demanding, therefore, further and detailed investigation. Updated amplitude analyses by other
authors \textit{\underline{including in the dataset}} all the high precision TOTEM measurements
at 7 TeV and 8 TeV can provide further checks on the results we have obtained and the 
conclusions we have drawn.

\section*{Acknowledgments}

We are grateful to an anonymous referee for valuable comments and 
suggestions, especially in respect to section 6.1. We are thankfull
to D.A. Fagundes and D.D. Chinellato for useful discussions.
Research supported by FAPESP (Contracts Nos. 11/15016-4, 09/50180-0).

\end{document}